\documentclass[twocolumn]{aastex63}
\usepackage{amsmath}

\DeclareMathAlphabet{\mathscr}{OT1}{pzc}%
                                 {m}{it}

\begin{document}

\title{Galaxy Clustering in the Mira-Titan Universe I: Emulators for the redshift space galaxy correlation function and galaxy-galaxy lensing}
\author{Juliana Kwan}
\affiliation{Astrophysics Research Institute, Liverpool John Moores University, 146 Brownlow Hill, Liverpool, L3 5RF, UK}
\author[0000-0002-6186-5476]{Shun Saito} 
\affiliation{Institute for Multi-messenger Astrophysics and Cosmology, Department of Physics, Missouri University of Science and Technology, 1315 N Pine St, Rolla, MO 65409, USA}
\affiliation{Kavli Institute for the Physics and Mathematics of the Universe (WPI), University of Tokyo, Kashiwa, Chiba 277-8583, Japan}
\author[0000-0002-3677-3617]{Alexie Leauthaud} 
\affiliation{Department of Astronomy and Astrophysics, University of California, Santa Cruz, CA 95064, USA}
\author[0000-0003-1468-8232]{Katrin Heitmann}
\affiliation{HEP Division,  Argonne National Laboratory, Lemont, IL 60439, USA}
\author{Salman Habib}
\affiliation{HEP Division,  Argonne National Laboratory, Lemont, IL 60439, USA}
\affiliation{CPS Division, Argonne National Laboratory, Lemont, IL 60439, USA}
\author[0000-0002-8469-4534]{Nicholas Frontiere}
\affiliation{HEP Division,  Argonne National Laboratory, Lemont, IL 60439, USA}
\affiliation{CPS Division,  Argonne National Laboratory, Lemont, IL 60439, USA}
\author[0000-0003-4936-8247]{Hong Guo}
\affiliation{Key Laboratory for Research in Galaxies and Cosmology, Shanghai Astronomical Observatory, Shanghai 200030, China}
\author{Song Huang}
\affiliation{Department of Astronomy and Tsinghua Center for Astrophysics, Tsinghua University, Beijing 100084, China}
\author{Adrian Pope}
\affiliation{CPS Division, Argonne National Laboratory, Lemont, IL 60439, USA}
\author{Sergio Rodrigu\'{e}z-Torres}
\affiliation{Departamento de F\'isica Te\'orica M8, Universidad Aut\'onoma de Madrid (UAM), Cantoblanco, E-28049, Madrid, Spain}

\begin{abstract}
We construct accurate emulators for the projected and redshift space galaxy correlation functions and excess surface density as measured by galaxy-galaxy lensing, based on halo occupation distribution modeling. Using the complete Mira-Titan suite of 111 $N$-body simulations, our emulators vary over eight cosmological parameters and include the effects of neutrino mass and dynamical dark energy. We demonstrate that our emulators are sufficiently accurate for the analysis of the Baryon Oscillation Spectroscopic Survey DR12 CMASS galaxy sample over the range $0.5 \leq r \leq 50\; h^{-1}$ Mpc. Furthermore, we show that our emulators are capable of recovering unbiased cosmological constraints from realistic mock catalogs over the same range. Our mock catalog tests show the efficacy of combining small-scale galaxy-galaxy lensing with redshift space clustering and
that we can constrain the growth rate and $\sigma_8$ to 7\% and 4.5\%, respectively, for a CMASS-like sample using only the measurements covered by our emulator. With the inclusion of a cosmic microwave background prior on $H_0$, this reduces to a 2\% measurement of the growth rate.
\end{abstract}


\section{Introduction}
Two-point clustering statistics, such as the two-point correlation function, form some of the most fundamental cosmological observables extracted from modern galaxy surveys. It is important to model these measurements as accurately as possible, including nonlinearities, such that the statistical power of the surveys can be fully exploited.  However, small-scale clustering is sensitive to the details of how galaxies occupy dark matter structure, gas physics, and nonlinear structure formation, all of which can be complicated to model. 

In particular, weak gravitational lensing and redshift space distortions, which both use forms of two-point statistics, are powerful, complementary tests of the growth of large-scale structure. 
Gravitational lensing involves the distortion of images of distant objects as light from background sources is perturbed by the gravitational potential of foreground structures. 
Weak lensing (WL) provides a wealth of information from the large-scale dark matter-dominated content of the Universe~\citep[cosmic shear, see, e.g.,][]{hikage19, heymans21, des22} to halo profiles and substructure~\citep[galaxy-galaxy lensing, see, e.g.][]{leauthaud17, singh20, lange21a, prat22} at small spatial scales. 
It is a primary target of many cosmological surveys, such as the Dark Energy Survey~\citep[DES;][]{flaugher15, des16, des18}, Hyper Suprime-Cam Survey~\citep[HSC;][]{aihara18, hikage19, hamana20}, the Kilo Degree Survey~\citep[KiDS;][]{kuijken15, heymans21}, the Vera C. Rubin Observatory Legacy Survey of Space and Time~\citep[LSST;][]{ivezic19} and the Nancy Grace Roman Space Telescope~\citep{spergel15}. 

Redshift space distortions provide additional information on the growth of large-scale structure as the peculiar motion of galaxies also traces the local gravitational potential. 
The measurement of redshift space distortions (RSD) is one of the few probes that is sensitive to the growth rate of large-scale structure, $f$, defined as $f = {d}D/{d \ln}a$, where $D$ is the linear growth factor and $a$ is the scale factor. 
Apart from being a useful quantity in itself, the growth rate also presents a means of distinguishing between different cosmologies that present the same cosmic expansion history but differ in terms of gravitational dynamics (e.g., modified gravity theories). 
Measuring RSD is a major component of every spectroscopic survey, including Baryon Oscillation Spectroscopic Survey~\citep[BOSS,][]{eisenstein11, dawson13}, SDSS-IV Extended Baryon Oscillation Spectroscopic Survey~\citep[eBOSS,][]{dawson16}, WiggleZ~\citep{drinkwater10}, FastSound~\citep{tonegawa15}, VIMOS Public Extragalactic Redshift Survey~\citep[VIPERS][]{guzzo14, scodeggio18} and 2dFGRS~\citep{colless01, cole05}. 
Indeed, satisfying the forecasted precision of many future probes, such as the Prime Focus Spectrograph~\citep{takada14}, Euclid~\citep{laureijs11}, the Hobby-Eberly Telescope Dark Energy Experiment~\citep{hill08} and the Dark Energy Spectroscopic Instrument~\citep[DESI,][]{desi16}, will rely on being able to obtain unbiased cosmological constraints from RSD. 

Currently, one of the limitations in our ability to extract information from these probes is the modeling of nonlinear structure formation. 
Typically this is done with expensive $N$-body simulations that can take many millions of CPU hours to achieve the appropriate level of accuracy. 
Furthermore, these upcoming surveys will require increasingly accurate models of the redshift space two-point clustering statistics if they are to achieve a 2\% measurement of the growth rate as intended~\citep{laureijs11, spergel13} and to avoid being dominated by systematic (as opposed to statistical) errors with increasing sky coverage and galaxy counts. 
Recent analyses, such as those by~\cite{chapman21, yuan22a, lange21b, zhai22}, have demonstrated the importance of using $\sim h^{-1}$Mpc scales for RSD by obtaining stronger constraints on $f\sigma_8$ than those using only large-scale observations.

In addition to requiring the nonlinear evolution of the dark matter clustering, it is also necessary to map the locations of galaxies within these structures, since these are our primary observables for most of the cases outlined above. 
Although there are techniques involving the use of hydrodynamics to incorporate gas physics into gravity-only simulations~\citep[see, for example][]{schaye15, mccarthy17}, these simulations cannot be done in a sufficient volume and quantity to be useful in cosmological analyses because of current limitations in computational power. 
Fortunately, galaxies can be placed into simulations in post-processing using a number of techniques 
that rely on modeling the distribution of galaxies within dark matter-dominated halos. These include~\citep[see][for a review]{wechsler18} halo occupation distribution~\citep[HOD; e.g.][]{berlind02}, Sub-Halo Abundance Matching~\citep[SHAM;
e.g.][]{kravtsov04, vale04, nuza13, saito16}, semi-analytic models of galaxy formation~\citep[SAMs;
e.g.][]{cole00, benson03}, and directly tracking halo substructure combined with halo merger tree information~(\citealt{jiang21, Sultan2021, korytov23}). 

The use of HOD modeling has been very successful as a proxy for inserting galaxies into $N$-body simulations and for cosmological parameter estimation (see for example~\citealt{vandenbosch13, cacciato13, more13a, mandelbaum13}). 
The method has also been widely applied to observational datasets~\citep{blake08, brown08, zheng09, white11, parejko13, guo15b}, since many of these two-point statistics can be calculated analytically via the halo model of clustering~\citep{cooray02, peacock00, seljak00, scoccimarro00, ma00, sheth01}. 
However, calibrating the halo model to reproduce the two-point clustering statistics to the required accuracy of current surveys necessitates direct input from $N$-body simulations because structure formation is difficult to model in the nonlinear, high density regime of halos as in the case of {\it Halofit}~\citep{smith03, takahashi12} and~\texttt{HMCODE}~\citep{mead15, mead16, mead21}. 

Emulation, and surrogate modeling in general, is an increasingly popular technique used to dramatically speed up the computation of summary statistics, such as the matter power spectrum~\citep{lawrence10, heitmann16,lawrence17, euclid19, euclid20,moran22}, the halo mass function~\citep{nishimichi19, mcclintock19, bocquet20}, galaxy two-point functions~\citep{zheng16, zhai19, lange19, wibking19, kobayashi20, wibking20, kokron21, zennaro21,lange21b, chapman21, zhai22, yuan22a, pellejeroibanez23} and higher order statistics~\citep{yuan22b, storeyfisher22} and also the likelihood function itself~\citep{pellejeroibanez20}.

Emulators are often implemented as a form of nonparametric regression facilitated by Gaussian processes (GPs) with the aim of reproducing highly nonlinear quantities, such as those measured from $N$-body simulations. 
The GP is trained on a number of carefully chosen simulations (applying some form of sampling theory) and is designed to return results with a given error tolerance within a prior parameter range. 
Instead of running a new simulation each time, the GP is conditioned such that its mean function (and variance) at a new set of parameters is consistent with the information contained in its input training functions.

In this paper, we present a set of new emulators that predict the galaxy projected correlation function, $w_p$, the monopole, $\xi_0^s$ and quadrupole, $\xi_2^s$ redshift space correlation functions, and the excess surface density, $\Delta\Sigma$, as a function of both HOD and cosmological parameters. 
We have obtained our nonlinear estimates of these quantities from the Mira-Titan suite of $N$-body simulations~\citep{heitmann16,bocquet20, moran22} as described in Section~\ref{sec:sims}. 
In Section~\ref{sec:HOD} we discuss our implementation of the seven-parameter HOD model used to produce mock galaxy catalogs.
Sections~\ref{sec:RSD}--\ref{sec:ggl} outline the theory behind the measurements and our underlying assumptions. 
Section~\ref{sec:emu} describes our technique for generating emulator predictions and the layout or design of training models employed. 
Our emulators are tested for the required accuracy in Section~\ref{sec:testing} using both out-of-sample and in-sample validation tests. 
In Section~\ref{sec:results}, we further demonstrate the ability of our emulators to recover unbiased cosmological constraints in a likelihood analysis using realistic mock galaxy catalogs. We also explore the constraints that might be attained from current datasets as a function of scale and using various combinations of probes. 
Section~\ref{sec:discussion} compares the performance of our emulators to others in the literature and discusses areas for improvement. 
We conclude in Section~\ref{sec:conclusions}.

Recently, there has been a surge of interest in the use of the nonlinear clustering of galaxies to constrain cosmological parameters, ~\citep[see, for example][]{chapman21, kokron21, salcedo22, yuan22a, contreras22,lange23,contreras23, zhai22}. 
Our emulators go beyond the current literature by combining both small-scale RSD and weak lensing that covers a wider range of cosmologies than those previously considered. 
By modeling the clustering of the CMASS sample of galaxies obtained from the twelfth data release of SDSS (BOSS DR12,~\citealt{reid16}) and galaxy-galaxy lensing from the \texttt{S16A} region of the HSC survey, we are also able to access a larger sample of galaxies with a higher signal to noise. 

The two-point statistics measured from CMASS galaxies are good candidates for modeling via emulation (or any other method for precision clustering observables) because they are exceptionally high signal to noise as a result of spectroscopic observations of $\sim$1.5 million galaxies covering nearly 10,000 deg$^2$.  
In addition, the CMASS galaxy catalog has been extensively studied for systematics~\citep{ross12, reid16}, widely used for cosmological parameter estimation~\citep[see for example][]{satpathy17}, and has publicly available data products\footnote{https://data.sdss.org/sas/dr12/boss/lss/} and mock catalogs~\citep{rodriguez16,kitaura16}. 
Furthermore, the CMASS footprint includes a substantial overlap with the HSC survey, which will supply the images required for galaxy-galaxy lensing.  
The HSC survey is designed to fully overlap with the CMASS footprint to facilitate joint analyses using both clustering and lensing.  
We will use images from the latest data release, known as \texttt{S16A}, which covers 137 deg$^2$ with a mean number density of $\sim$22 galaxies per square arcminute~\citep{aihara18}. 

A subset of CMASS galaxies has been analysed by~\citet{zhai22} using emulators for $w_p$, $\xi_0^s$ and $\xi_2^s$ but they found a lower $f\sigma_8$ value than that predicted by~\citet{planck18} and a nonzero detection of effects beyond General Relativity (GR) for their highest redshift sample. 
Furthermore, their analysis did not include the additional information contained from cross correlations with galaxy-galaxy lensing observations using the same sample.  

An offset of 20--30\% in the large-scale bias between $\Delta\Sigma$ and $w_p$ measured from the same sample of galaxies was first observed by~\citet{leauthaud17} under the assumption of a Planck best-fitting cosmology. 
This has been subsequently confirmed in other studies with different samples~\citep[e.g. BOSS LOWZ;][]{lange21a}.

One advantage of our approach compared to previous analyses involving the CMASS sample, ~\citep[e.g][]{chapman21, yuan22a, zhai22}, is that we model galaxy-galaxy lensing in conjunction with redshift space clustering, which is important in breaking parameter degeneracies and constraining $\sigma_8$ and H$_0$ in particular. 
It would also be useful to analyse a different galaxy sample from~\citealt{lange23}, such as CMASS, to confirm the recent trend of measuring lower values of $\sigma_8$ and $f$ from low redshift nonlinear clustering analyses than those preferred by Planck.

\section{Simulations}\label{sec:sims}
We use the Mira-Titan Universe suite of $N$-body simulations to generate training samples for our emulators.
The simulation suite has been specially designed for this purpose and has been used previously for emulating the density fluctuation power spectrum~\citep{lawrence17, moran22} and the halo mass function~\citep{bocquet20}. The suite consists of 111 models covering cosmologies with dynamical dark energy and massive neutrinos over the following parameter range:
\begin{eqnarray}
   0.12\le&\omega_{\rm cdm}&\le 0.155,\\
   0.0215\le&\omega_{\rm b}&\le 0.0235,\\ 
   0.7\le&\sigma_8&\le 0.9,\\ 
   0.55\le& h&\le 0.85,\\
   0.85\le& n_s& \le 1.05,\\ 
   -1.3\le &w_0 & \le -0.7,\\
   -1.73\le &w_a & \le 1.28,\\
   0.0\le&\omega_\nu & \le 0.01.
\end{eqnarray}
The $w_0, w_a$ parameterization~\citep{chevalier01, linder03} is used to characterize the time variation of the equation of state parameter of dark energy as $w = w_0 + w_a (1-a)$, where $w_0$ and $w_a$ are the parameters varied in the simulations. 
Note that the values of $w_0$ and $w_a$ have an additional constraint, that $(w_0 + w_a) < 0 $, to avoid dark energy domination during the early universe.

Briefly, the models of the Mira-Titan Universe were chosen to satisfy a (sequentially convergent) eight-dimensional space-filling design, as described in~\citet{heitmann16}, to obtain percent-level accuracy predictions for the nonlinear matter power spectrum, valid for massive neutrino and dynamical dark energy models. 
Note that the uniform nature of the sampling process allows for unbiased error controls throughout the parameter space. 
This is not the case for emulators built with a nonuniform sampling prior (such as those preferentially centered around a Planck cosmology).
Each cosmology is represented by a single high-resolution $N$-body simulation with a box length of 2100 Mpc and containing 3200$^3$ particles with a force resolution of $\sim$6.6 kpc, and a particle mass resolution of $\sim1.177 \times 10^{10} ({\Omega_m h^2}/{0.15})$ M$_{\odot}$. 
The simulations were run using Hardware/Hybrid Accelerated Cosmology Code~\citep{habib16} on the Mira and Titan supercomputers. 
The simulations were initialized at $z=200$ using the Zel'dovich approximation. 
The accuracy of our neutrino implementation (neutrinos are not included as a separate species but instead are modeled in the evolution of the background cosmology) is discussed in~\citet{upadhye14}. 
This approximation is accurate to 1\% in the matter power spectrum when compared to simulations that treat neutrinos as separate particles for the range of light neutrino masses that are spanned by the emulator~\citep{castorina15, sullivan23}.

Halos are identified using a friends-of-friends (FOF) algorithm with a linking length of $b=0.168$ with a minimum number of 40 particles. 
This is motivated by previous HOD studies in the literature, such as~\cite{white11} and \cite{guo15}, in which a lower value of $b$ (as opposed to the more conventional value of $b=0.2$) is preferred to produce a more compact halo and reduce spurious linkages. 
In addition, the calculation of the central halo velocity is less likely to be skewed by particles on the outskirts (see Section~\ref{sec:velbias}). 

The emulators are built using a single Mira-Titan snapshot at $z = 0.539$, rather modeling the full redshift range of the CMASS catalog from $0.43 < z < 0.7$. 
The HOD of the CMASS galaxies varies slowly enough, and moreover, the redshift distribution is sufficiently peaked such that \cite{saito16} found this approximation to affect the clustering by less than a percent.
We have also neglected to account for the footprint of the DR12 survey or fiber collisions, preferring instead to treat these in the observations (or mocks) via the Landy-Szalay estimator~\citep{landyszalay93} and the fiber collision corrections presented in~\citet{guo12} respectively.

\subsection{HOD Modeling}\label{sec:HOD}
We use an HOD model to insert synthetic galaxies into the $N$-body simulations. 
In HOD modeling, the number of galaxies assigned to each halo is solely determined by the mass of the halo. 
Generally, there are two mass thresholds in the model: the first must be satisfied for the halo to host a central galaxy and the second, at a higher mass, allows for the possibility of hosting additional satellite galaxies. 
The parameters of the model are tuned to reproduce observational characteristics, e.g. the low satellite fraction of CMASS can be achieved by setting a higher mass threshold for the satellites. 
While baryonic effects are expected to make a contribution to small-scale clustering, e.g. the presence of AGN feedback causes suppression in the total matter power spectrum, there is some evidence to suggest that the effect is much smaller in redshift space for our scales of interest~\citep{hellwing16}.

Our HOD model is based on the work in~\citet{zheng05}, which has been successfully applied to a 
number of studies involving CMASS galaxies~\citep{white11, guo14, reid14, guo15,zhai22}. 
As in some of these previous studies, we include an additional free parameter, $f_{\rm max}$, to account for sample incompleteness by limiting the maximum probability that any halo can host a galaxy as well as free parameters for the velocity bias.

For the velocity bias, we follow the~\citet{guo15} and~\citet{reid14} approach in spirit but our halo masses have been defined differently. 
In this model, the mean number of central galaxies $\langle N_{\rm cen} \rangle$ contained in each halo of mass, $M$, and the number density of centrals, $\bar{n}_{\rm cen}$ in the full sample is given by
\begin{eqnarray}
\langle N_{\rm cen}(M) \rangle &=& \frac{f_{\rm max}}{2} \left [ 1 + {\rm erf} \left( \frac{\log M - \log M_{\rm cut}}{\sigma}\right)\right],\\
\bar{n}_{\rm cen} &=& \int \langle N_{\rm cen}(M) \rangle \frac{dn}{dM} dM, 
\end{eqnarray}
where $M_{\rm cut}$ is the mass threshold for central galaxies, $\sigma$ is the spread in the mass cut, $\frac{dn}{dM}$ is the halo mass function and $f_{\rm max}$ is the halo incompleteness factor, which allows for missed central galaxies in the observational sample because of a luminosity or color cutoff. 
The mean number of satellite galaxies contained in a halo of mass, $M$, and the 
corresponding mean number density of satellites, $\bar{n}_{\rm sat}$, is given by
\begin{eqnarray}
\langle N_{\rm sat}(M) \rangle &=& \frac{\langle N_{\rm cen} \rangle}{f_{\rm max}} \left(\frac{M-\kappa M_{\rm cut}}{M_1}\right)^\alpha, \\
\bar{n}_{\rm sat} &=& \int \langle N_{\rm sat} \rangle \frac{dn}{dM} dM, 
\end{eqnarray}
where $\kappa$ modulates the fraction of halos containing centrals that also host satellites, $M_1$ is approximately the mass required for the first satellite galaxy and $\alpha$ controls how many satellite galaxies can be hosted by a single halo. 
Central galaxies are chosen from a Bernoulli distribution with the probability $\langle N_{\rm cen} \rangle$ while the number of satellite galaxies is assumed to be sampled from a Poisson distribution at a rate of $\langle N_{\rm sat} \rangle$. 
The central galaxy is given the position and velocity of the center of the halo, plus an additional
term to represent an offset between the velocity of the halo center and the galaxy (known as the velocity bias; we will elaborate on this further in Section~\ref{sec:velbias}). 
We define the halo center as  the minimum of the gravitational potential of the halo and the velocity of the center is calculated by averaging the velocities of all particles belonging to the FOF halo. 
This is in contrast to~\citet{reid14} and~\citet{guo15}, who only used a fraction of the innermost particles to define the central velocities.  
The coordinates and velocities of satellite galaxies are assigned to $\langle N_{\rm sat} \rangle$ dark matter particles, chosen at random from the halo and may also experience a velocity bias, defined as a difference between the velocity of the halo particles and velocity of the halo center (see also Section~\ref{sec:velbias} for details). 

Within our emulator, we vary the five HOD parameters: $\sigma$, $\alpha$, f$_{\rm max}$, $\bar{n}$, the number density of galaxies, and f$_{\rm sat}$, the satellite to central fraction, and two velocity bias parameters, $\alpha_c$ and $\alpha_s$ for the redshift space multipoles. 
These parameters cover the following ranges: 
\begin{eqnarray}
   0.01\le&\sigma&\le 0.6,\\
   0.7\le&\alpha&\le 1.5,\\ 
   0.7\le& f_{\rm max}&\le 1,\\ 
   -3.7 \le& \log_{10} \bar{n}/(h^3{\rm Mpc}^{-3}) &\le -3.2,\\
   0.05\le& f_{\rm sat} & \le 0.15,\\ 
   0.0\le& \alpha_c & \le 1.5,\\ 
   0.0\le & \alpha_s & \le 1.5.
   \label{eqn:HOD_parameters_space}
\end{eqnarray}
These ranges are chosen to bracket the observed clustering from the CMASS DR12 sample of galaxies. The parameters are varied in addition to the eight cosmological parameters covered within the Mira-Titan Universe suite. 
The HOD parameters have been chosen to match the published parameter values for CMASS~\citep{white11, reid14} and also to maximize the available range in halo masses based on the resolution of our simulations. 
We have replaced the usual halo mass threshold parameters, $M_{\rm cut}$ and $M_1$ for central and satellite galaxies, respectively, with $\bar{n}$ and f$_{\rm sat}$. 
This is because we have strong prior information on the latter from previous analyses involving CMASS galaxies  (see~\citealt{white11, reid14}). 
We found that varying $\kappa$ has little effect on the final results, so we do not consider it as a free parameter~\citep{kwan15} and assume $\kappa=1$ throughout the paper. 
The velocity bias parameters, $\alpha_c$ and $\alpha_s$, defined in Section~\ref{sec:velbias} do not present a clear, intuitive range for coverage by the emulator, but there is evidence to suggest that massive red galaxies are moving at some fraction of the dark matter content~\citep[see, for example,][]{guo15, yuan21b, yuan22c}. 
These last two parameters are omitted when modeling galaxy-galaxy lensing. 

Since imposing an HOD model involves a degree of stochasticity as to the exact centrals and satellites chosen, in terms of both location and number, we average over 20 realisations for each measurement of $\xi(s,\mu)$ and $w_p$. 
We found that this was sufficient to produce less than 1\% difference between any given realization and the final, averaged quantity. 
Being of lower signal-to-noise in the observations, the $\Delta\Sigma$ emulator only required a single realization to satisfy the observational error requirement. 

\subsection{Redshift Space Distortions}\label{sec:RSD}
In our $N$-body simulations, we convert the real space positions to redshift space by perturbing
each particle position using the peculiar velocity along the line of sight, $u_x$, as follows: 
\begin{equation}
s = x + \frac{u_x (1+z)}{H(z)},
\label{eqn:particleshift}
\end{equation}
where $x$ and $u_x$ are the coordinate and velocity along the line of sight, respectively, $z$ is the redshift, and $H(z)$ is the Hubble relation. 
We use the plane parallel approximation, i.e., our line of sight is taken to be along one axis of the simulation volume, when applying Equation~\ref{eqn:particleshift} to our mock galaxies. 
The distant observer approximation holds if the angle separating the galaxies is small. Since the greatest pair separation we consider is 50 $h^{-1}$Mpc at a distance of $z\sim0.55$ (the mean of the CMASS redshift distribution), this should not pose a significant source of error for our analysis.
We use the Landy-Szalay estimator~\citep{landyszalay93} in redshift space to measure the 2D correlation function, $\xi(s,\mu)$, as follows: 
\begin{eqnarray}
&&\xi^s_{gg}({\Delta s},{\Delta \mu}) = \nonumber\\
&&\frac{DD({\Delta s},{\Delta \mu})- 2DR({ \Delta s},{\Delta \mu}) + RR({\Delta s},{\Delta \mu})}{RR({ \Delta s},{\Delta \mu})}, 
\end{eqnarray}
where $DD$, $DR$, and $RR$, represent the counts of data-data, data-random, and random-random pairs within a bin with widths $\Delta s$ and $\Delta \mu$ respectively.
Then $\xi(s,\mu)$ can be decomposed into multipole moments as follows:
\begin{equation}
\xi^s_\ell \equiv \frac{2\ell+1}{2}\int^{+1}_{-1} d\mu \; \xi^s(s,\mu)\; L_\ell(\mu),
\label{eqn:multipole}
\end{equation}
where $\mu = \cos(\bf{u} \cdot \bf{r})$ represents the angle between the velocity, $u$, with respect to the line-of-sight vector, $r$ and $L_\ell(\mu)$ are the Legendre polynomials; $L_0(\mu) = 1$, $L_2(\mu) = (3\mu^2-1)/2$ for the monopole and quadrupole, respectively. 
We measure $\xi(s,\mu)$ using~\texttt{corrfunc}~\citep{sinha19,sinha20} from the Mira-Titan simulations at $z=0.539$ (near the peak of the CMASS redshift distribution) after creating the mock HOD catalogs and applying the transformation in Equation~\ref{eqn:particleshift}. 
We have used 21 bins in $s$ between $0.1 \le s \le 50$ Mpc spaced logarithmically and then an additional nine bins linearly spaced between $50 < s < 100$ Mpc. 
The angular spacing is initially $\Delta\mu = 0.005$ when performing the measurements with~\texttt{corrfunc} and then we average over several bins such that $\Delta\mu = 0.02$ when calculating the multipole moments. 

\subsubsection{Velocity Bias}\label{sec:velbias}
The velocity bias is intended to model any offsets between the velocity distribution of the dark matter particles and the galaxies occupying the halo. 
There are two velocity biases to consider: a velocity bias between the halo center and the central galaxy, $\alpha_c$, and the intrahalo velocity bias, $\alpha_s$, between the dark matter halo particles and the satellite galaxies. 
These could be caused by the movement of the central galaxies within the host halo, since the relaxation of the dark matter halo does not ensure that the central galaxy is also at rest, and/or an offset between the velocity distributions of the galaxies and the dark matter in the host halo. 
A further complication is that the assignment of velocities to HOD galaxies is not uniquely determined, because the velocity of the halo center depends upon the assumption of a core radius, or radial extent taken to be the edge of the halo and this can affect the significance of the detection of a velocity bias in the observed redshift space clustering. 
Indeed, there are strong indications that the halo bulk velocity calculated using the virial radius, $R_{\rm vir}$, is offset from the halo central velocity defined within a fraction (say 0.1-0.2) of $R_{\rm vir}$~\citep{behroozi13}. 
This arises because the efficiency in dynamical friction is higher at the halo outskirts (i.e. significant momentum transfer between the halo and incoming satellites) and then decreases near the halo core.

Here we discuss the most common methods presented in the literature and compare these with our own approach for clarity. 
Because of the result in~\citet{behroozi13}, previous work in the literature used a core radius for the calculation of the central halo velocity that is a fraction of $R_{\rm vir}$. 
\citet{reid14} used a spherical overdensity (SOD) algorithm to identify halo centers from peaks in the density distribution. 
The central velocity was calculated by averaging the velocities of the innermost 20 particles belonging to the halo or $\sim$3.7\% of all halo member particles. 
This is the velocity assigned to the central galaxy as well as a velocity bias that is proportional to the virial velocity of the halo. 
They conclude that there is no clear evidence for a central velocity bias, although they were unable to vary the velocity biases freely because of the additional computational cost involved with the~\cite{neistein12} precompute method.

\citet{guo15} used both FOF catalogs with $b=0.17$ and SOD catalogs for their analysis.
However, unlike~\citet{reid14}, the velocity of the halo center and velocity dispersion is defined from the innermost 25\% particles. 
This resulted in a much stronger conclusion for the detection of a central velocity bias than reported by~\cite{reid14} but ultimately the significance is lessened when one accounts for the differences in their definition of the velocity of the central galaxy. 
More recently,~\citet{lange21b} used the inner 10\% of halo particles to compute a core velocity.  
In our simulations, halos are identified by the FOF algorithm with $b=0.168$ and the central velocity and velocity dispersion of each halo are calculated using all the linked particles. 
Since there is no consensus in the literature or a physical justification for the actual value of the radius used to calculate the central halo velocities, we have opted for the minimal assumption of imposing no radial cut. 
In compensation, we have instead constructed our emulators over a generous range on both velocity bias parameters, since we consider these to be nuisance parameters and are ultimately only interested in determining the cosmology. 
That is, we use the complete set of particles belonging to an FOF halo to calculate both its dispersion and central velocity and allow the velocity biases to absorb any residual differences. 
This is by no means a definitive choice and we leave the definition of an appropriate radius to future work. 
Our implementation of both velocity bias parameters, $\alpha_c$ and $\alpha_s$, is as follows: 
\begin{eqnarray}
    {\rm u_{r; \, cen}} &\rightarrow&  {\alpha_c} \sigma_v + {\rm u_{r; \, cen}},   \\
    {\rm u_{r; \, sat}} &\rightarrow&  {\alpha_s} ({\rm u_{r \; sat}} - {\rm u_{r; \, cen}}) + {\rm u_{r; \, cen}},
    \label{eqn:velocitybias}
\end{eqnarray}
where u$_{\rm r; \, cen}$ (u$_{\rm r; \, sat }$) is the velocity of the central (satellite) galaxy, and $\sigma_v$ is the 1D velocity dispersion calculated from all the particles contained in the halo. 
This is similar to the~\cite{reid14},~\cite{guo15} and~\citet{lange21b} definitions, except their measurement of the halo central velocities occurs within a much more compact radius. 
This is important to bear in mind when comparing our results on the velocity bias parameters to those in the literature, as our conclusions are likely to differ on the basis of our definition of the velocity bias. 

We have also included a redshift measurement error individually for each mock galaxy, in which an additional Gaussian velocity dispersion centered on zero with a width of 0.44 $h^{-1}$Mpc is allowed to perturb its position along the line of sight.
This was shown to be necessary for the CMASS sample in~\citet{reid14} and~\citet{guo15} and we apply this correction to both the measurements that are used to construct the emulators and to the realistic mock galaxy catalogs that will serve as a test of their accuracy in a later section.  

\subsection{Projected Correlation Function}\label{sec:wp}
The projected correlation function is defined as
\begin{equation}
    w_p(r) = 2\int^{\infty}_0 \xi^s_{gg}(s_\parallel,s_\perp) \; ds_{\parallel},
    \label{eqn:wp}
\end{equation}
where $\xi_{gg}$ is the redshift space 2D correlation function between two galaxies. 
Within a finite volume, this is done by integrating the pair counts in redshift space to a maximum parallel separation, $s_{\parallel; \rm max}$ which we take to be 150 Mpc. 
We use the same value of $s_{\parallel; \rm max}$ for calculating the observed and mock projected correlation functions to which we will be comparing our emulator against throughout this work.
We use the same bin spacing and range of scales for $w_p$ as for the redshift space multipoles and measure using the {\tt corrfunc} package. 
\citet{reid14} found that it was important to include $w_p$ when jointly fitting for the HOD parameters from the redshift space monopole and quadrupole moments since small-scale measurements of $w_p$ can be quite informative about the internal structure of the halo and aid in constraining the distribution of satellite galaxies.

\subsection{Galaxy-Galaxy Lensing}\label{sec:ggl}
Galaxy-galaxy lensing occurs when the dark matter halo of the galaxy sample in consideration (the lens galaxies) distorts the images of background or source galaxies. 
On small scales, it provides an additional source of information about the halo substructure and how galaxies populate halos. 

Combining weak lensing and clustering can break many parameter degeneracies, such as that between the amplitude of the dark matter power spectrum and the galaxy bias, namely, $\sigma_8$ and $b$, since lensing is directly sensitive to the halo mass, and furthermore, each probe is subject to its own set of systematic uncertainties.

The projected surface density, $\Sigma$ for a transverse distance, $r$, is defined as:
\begin{equation}
    \Sigma(r) = 2 \, \bar{\rho}_m \int^\infty_r 
    \left[ 1+\xi_{\rm dm,g} \right] \frac{R \; dR}{\sqrt{R^2-r^2}},
    \label{eqn:sigmadef}
\end{equation}
where $R^2 = r^2 + {\pi}^2$, with $\pi$ as the radial distance, and $\xi_{\rm dm,g}$ is the galaxy-dark matter cross-correlation function.
For the light-cone mocks used in Section~\ref{sec:results}, we use the following estimator for the measurement of $\xi_{\rm dm,g}$:
\begin{eqnarray}
&&\xi_{gm}({\bf \Delta r}) = \nonumber\\
   &&\frac{D_gD_m({\bf \Delta r}) - D_gR({\bf \Delta r}) - D_mR({\bf \Delta r}) + RR({\bf \Delta r})}{RR({\bf \Delta r})},\nonumber\\
   \label{eqn:ls-cross}
\end{eqnarray}
where $D_g$ represents the mock galaxies and $D_m$ the dark matter particles in the same light-cone.

Galaxy-galaxy lensing measures the surface overdensity, $\Delta\Sigma$, the observable of interest, as:
\begin{eqnarray}
    \Delta\Sigma(r) &=& \bar{\Sigma}(<r) - \Sigma(r) \label{eqn:deltasigmadef}\nonumber\\
    &=& \Sigma_{\rm crit}\gamma_t(r), 
\end{eqnarray}
in comoving coordinates, where $\Sigma_{\rm crit}$ is the critical surface density defined as
\begin{equation}
    \Sigma_{\rm crit} = \frac{c^2}{4\pi G {(1+z_l)}^2} \frac{D(z_s)}{D(z_l,z_s) D(z_l)},
\end{equation}
with $D(z_{l/s})$ is the angular diameter distance to the lens/source redshift and $\gamma_t$ is the tangential shear. 
Note that we consistently use comoving coordinates for $\Delta\Sigma$ throughout this paper. 
We model $\Delta\Sigma$ using a single snapshot at $z=0.539$ by measuring the cross-correlation function between a 1\% downsampled population of $N$-body simulation particles (as the full particle snapshots have not been saved) and mock HOD catalogs at that redshift. 
We then use the \texttt{Halotools} package~\citep{hearin17} to calculate $\Delta\Sigma(r)$ from the Mira-Titan simulations. 
For our light-cone mocks, we used Equation~\ref{eqn:ls-cross} as measured by \texttt{corrfunc} and then converted this into $\Delta\Sigma$ using Equations~\ref{eqn:sigmadef} and~\ref{eqn:deltasigmadef} since \texttt{Halotools} only operates on simulation volumes.
We have tested that there is no appreciable difference between using the downsampled and full particle snapshots to percent-level precision as tested against our previous $\Delta\Sigma$ emulator~\citep{kwan15}.

\section{Emulation}\label{sec:emu}
We now describe the emulation methodology used to make fast and accurate predictions of the nonlinear galaxy projected correlation function, monopole and quadrupole redshift space correlation functions, and the projected surface density. 
Emulation combines Bayesian statistical methods with machine learning, providing a way of predicting smoothly varying functions from a set of precalculated models, by assuming that the function space can be well described by a Gaussian Process~\citep[for a review, see][]{rasmussen06}. 
We refer the interested reader to~\cite{heitmann09} for further details on the emulation method within a cosmological context. 
Unlike fitting functions, the errors are well defined across a predetermined range of parameter values using in-sample and out-of-sample validation methods. 

\subsection{Design}\label{sec:design}
The emulator needs to be trained on an initial set of models to determine the hyperparameters that govern the GP. 
Generally, it is important to choose a design strategy that fills the allowed parameter range with the fewest number of models since each experiment is computationally expensive. 
This is achieved in the Mira-Titan Universe by using a space-filling design similar to a tessellation~\citep{heitmann16}. 
Since the cosmological part of the parameter space has already been chosen, it only remains to determine a sampling strategy for the HOD models for which we have adopted an Optimized Latin Hypercube. 
Each $N$-body simulation uses the same hypercube design for the HOD parameter space, since this is necessary for our method of constructing the emulators, which takes advantage of the separability of the design, which we discuss below.

\subsection{GPs and Large Data Sets}\label{sec:BCM}
The standard Gaussian Process scales as $\cal{O}$$(N^3)$, where $N$ is the number of design points because each evaluation involves performing a matrix inversion. 
It is impractical, therefore, to use this method for building an emulator for our particular problem, because of the large number of design points involved, since we not only have the 111 cosmologies of the Mira-Titan suite, but also each simulation must give rise to a number of HOD models to cover the range of CMASS galaxy clustering. 

There are a few methods that can deal with this situation, such as sparse GPs, and Bayesian Committee Machines~\citep[BCMs;][]{tresp00, deisenroth15}. 
We have chosen to use a Kronecker structured design, as developed in~\citet{saatci11} and applied to cosmology in~\citet{yuan22a}, and we leave the exploration of the efficacy of some of these other methods in Appendix A. 
We use the Python package, GPy\footnote{http://sheffieldml.github.io/GPy/}, to compute the GPs. 

The Kronecker GP requires that the covariance is separable such that 
\begin{equation}
    K = \bigotimes^{D}_{d=i} K_d,
\end{equation}
where $K$, the total covariance matrix is composed of the Kronecker product of $D$ individual covariances matrices, $K_d$. 
In our case, we can see that this condition may be fulfilled by taking $D=2$ and allocating a separate $K_d$ for the cosmology and the HOD components.
Note that this is possible because the design space is easily separable into cosmology and HOD components, since the HOD models are applied on top of each simulation, which themselves follow a design independently of the HOD parameter space. This does not imply that the GP is treating the HOD and cosmology separately, which we will address in Appendix A.
We use a radial basis function (RBF) or squared exponential kernel for each $K_d$ as given by 
\begin{equation}
    K_d({\bf x_i}, {\bf x_j}) = \sigma^2 \exp \left[- \frac{1}{2l^2} ({\bf x_i}, {\bf x_j)}^T({\bf x_i}, {\bf x_j}) \right],
    \label{eqn:RBF_kernel}
\end{equation}
where $\sigma$ and $l$ are hyperparameters to be determined by the training sample and ${\bf x_i}$ is a vector representing $i$ design parameters. 
We also add a white noise term to the kernels to account for the fact that our measurements are not perfect representations of each two-point statistic. In addition, GPs often require a small amount of jitter in the form of white noise to avoid over-constraining the problem.
Because the full kernel is separable into smaller parts, the GP scales as $\cal{O}$(N $\times$ M) for $N$ cosmologies and $M$ HOD models. 
For each cosmology, we measure $w_p, \xi_0^s, \xi_2^s$ (and $\Delta\Sigma$) from 600 (150) HOD models in a Latin Hypercube design spanning seven (five) parameters as described in Equation~\ref{eqn:HOD_parameters_space}. 
We then build a basic GP using only an RBF kernel to make predictions for these observables from our set of HOD parameters for each of our 111 cosmologies. 
We have confirmed using out-of-sample HOD predictions that these GPs are accurate at the percent level for each cosmology. 
These emulators are used to make predictions of the same 200 HOD models (1000 for the quadrupole) for each cosmology that are fed into the Kronecker GP for hyperparameter optimization. 
The reason an initial emulator is required is that the HOD models do not use $\bar{n}$ and $f_{\rm sat}$ as input parameters, rather these are derived parameters to be measured after the catalog is made from specifying $M_{\rm cut}$ and $M_{1}$. 
This ensures that each set of HOD models is the same regardless of cosmology which is a necessary design trait for the Kronecker GP. 
The number of design points was determined empirically for each probe and we found that while only 200 models were sufficient for every statistic, the performance of the emulator for the quadrupole moment continues to improve up to 1000 models because of the large variance in values around the region of $s$ values where the quadrupole moment changes sign.
We have not determined if additional models would further improve the accuracy of the quadrupole emulator since adding more models would slow the predictions.
We test the accuracy of our emulators in the following section.

\section{Emulator Testing}
\label{sec:testing}
We explore the uncertainty in our emulator predictions by performing a series of in-sample and out-of-sample validation tests. 
The out-of-sample validation tests give the closest approximation to the true error of the emulators, since we are both using the complete design and these models are external to the original training set. 
However, these tests are expensive since they require a new $N$-body simulation for each new prediction.
In-sample (or holdout) tests involve removing one design point from each emulator and then using the remaining models to predict that point. 
This overestimates the error since a smaller design is used and excluded models on the design edge become beyond the scope of the test emulator. 
We use a mixture of in-sample and out-of-sample validation tests to ensure that our emulators are sufficiently accurate. 
Our error estimates show that our $w_p$ and $\Delta\Sigma$ emulators are accurate to at least $\sim$2\% precision over the full range of 0.1 to 50 $h^{-1}$Mpc. 
The error on our monopole and quadrupole emulators varies between 2--5\%, reaching the latter only on sub-$h^{-1}$Mpc and $>10 h^{-1}$Mpc scales.

\begin{figure}[t]
    \centering
    \includegraphics[width=\linewidth]{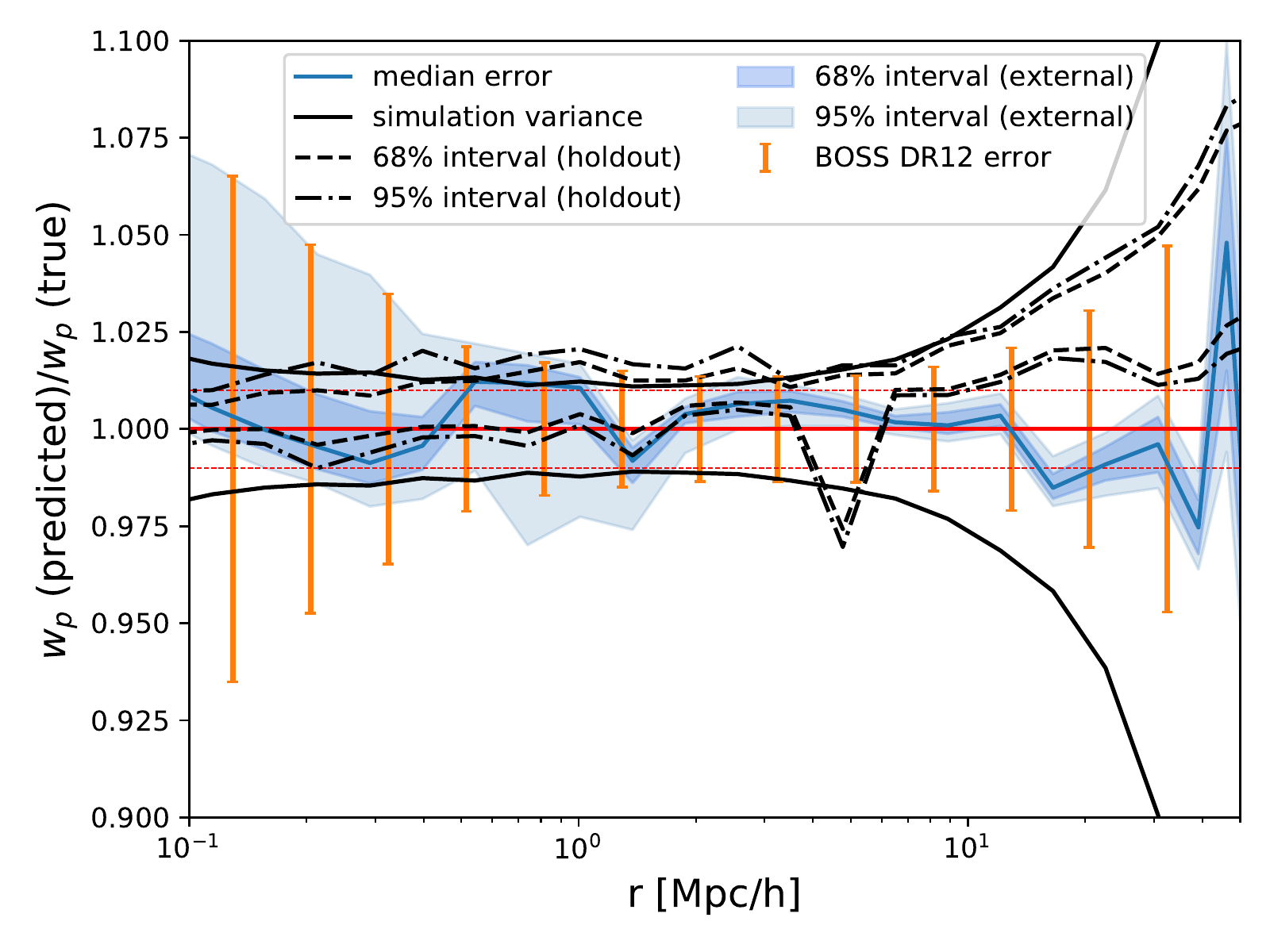}
    \caption{Accuracy test of the $w_p$ emulator using both out-of-sample and holdout methods. For the out-of-sample test, an external cosmology is chosen and fully simulated. 150 HOD models are then generated according to a seven parameter symmetric Latin Hypercube (LHS) design using this simulation and compared against our emulator. The blue curve is the median error in the prediction and the shaded regions show the 68\% and 95\% confidence intervals.
    For the holdout tests, with confidence intervals represented in dashed (68\%) and dotted-dashed (95\%) lines, we exclude one of five central cosmologies in the 111-model design at a time while the remaining models are used to construct a new emulator to predict the missing model.
    The jackknife variance estimated from a single simulation volume is shown in solid black. 
    We have also shown the 1-$\sigma$ error bars of $w_p$ measured from the CMASS DR12 sample
    using jackknife resampling in orange.}
    \label{fig:wp_accuracy_test}
\end{figure}

\begin{figure}[t]
    \centering
    \includegraphics[width=\linewidth]{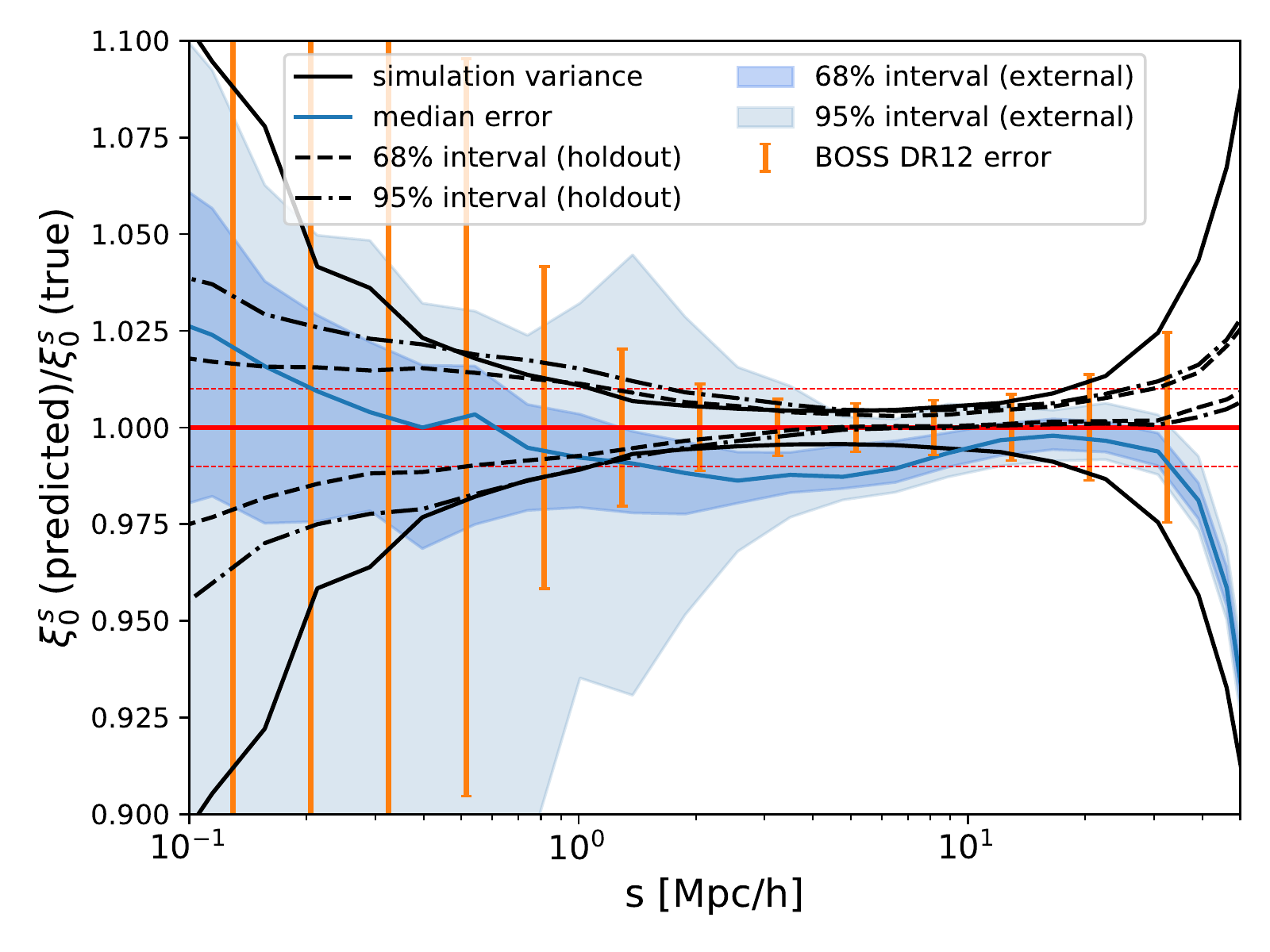}
    \caption{Out-of-sample and holdout accuracy tests of the $\xi_0^s$ emulator. 
    Similar to Figure~\ref{fig:wp_accuracy_test}, we use 150 HOD models for the external test laid in an optimized seven-parameter symmetric Latin Hypercube (LHS) design from the same external cosmology to measure the ratio between these models and the predicted monopole moment from the emulator.  
    The blue curve is the median error in the prediction and the shaded regions show 
    the 68\% and 95\% confidence intervals from the out-of-sample test. 
    The same five internal cosmologies as in Figure~\ref{fig:wp_accuracy_test} have been excluded for the holdout tests and their confidence intervals are shown in black dashed (68\%) and dotted-dashed (95\%) lines. 
    Again, the simulation variance is shown in solid black. 
    For comparison, we also show the 1-$\sigma$ error bars of $\xi_0^s$ measured from the CMASS DR12 sample using jackknife resampling.}
    \label{fig:xi_mono_accuracy_test}
\end{figure}

\subsection{Out-of-sample Validation Tests}\label{sec:m000}
Our out-of-sample validation tests are performed using a reference Mira-Titan simulation with a WMAP7-like cosmology: $\Omega_m$ = 0.2648, $\Omega_b$ =0.04479, $h_0$ = 0.71, $\sigma_8$ = 0.8, $n_s$ = 0.963, $w_0$ = -1, $w_a$ = 0, and $\Omega_\nu$ = 0. 
This is chosen to be near the center of the cosmological design space. 
This simulation has a box length of 2100 Mpc and 3200$^3$ particles, giving rise to a mass resolution of $\sim$ 7.43 $\times$10$^9$ $h^{-1}$M$_{\odot}$. 
For consistency, this run is produced in exactly the same manner as the other Mira-Titan simulations used to build the emulators. 
We chose to generate our test HOD models in this cosmology according to another Latin Hypercube, to ensure that the HOD design space is evenly sampled. 
Each HOD model is also the average of 20 realizations to prevent errors from scatter. 

\begin{figure}[t]
    \centering
    \includegraphics[width=\linewidth]{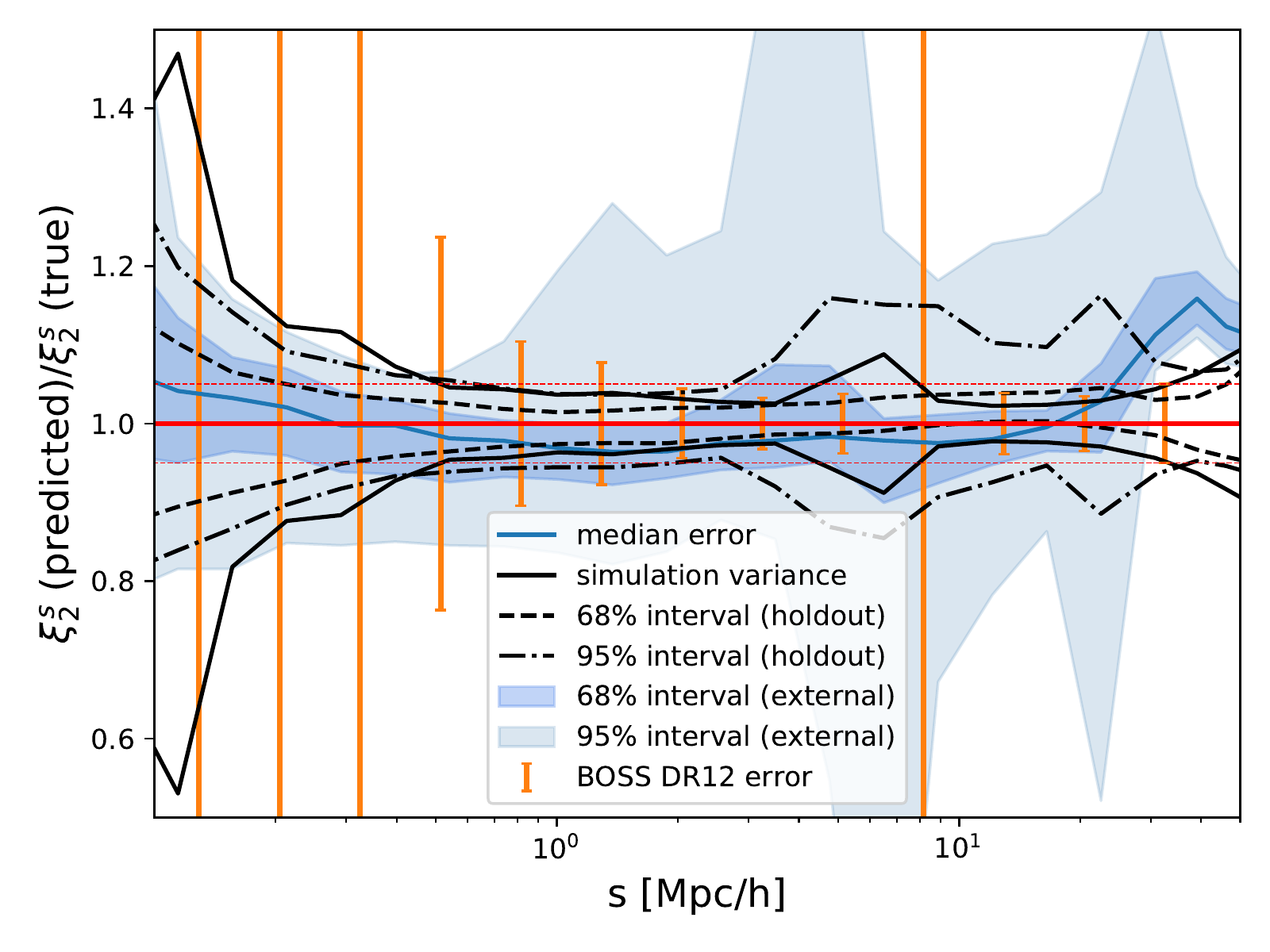}
    \caption{Accuracy test of the $\xi_2^s$ emulator. As in previous figures, we compare the predictions of the emulator against 150 HOD models selected from an external cosmology and show the 68 and 95 percent confidence intervals in the blue shaded regions. Again the errors from the five holdout cosmologies are in black dashed (68\%) and dotted-dashed (95\%) lines and the simulation measurement error is in solid black. 
    For comparison, we also show the 1-$\sigma$ error bars of 
    $\xi_2^s$ measured from the CMASS DR12 sample using jackknife resampling.
    Because the quadrupole moment crosses through zero on the scales considered, plotting the ratio gives a misleading estimate of the error when both the prediction and the truth are small. 
    This accounts for the sharp increase in errors near $\sim$5 $h^{-1}$Mpc scale in both 
    the emulator and observations.}
    \label{fig:xi_quad_accuracy_test}
\end{figure}

\begin{figure}[t]
    \centering
    \includegraphics[width=\linewidth]{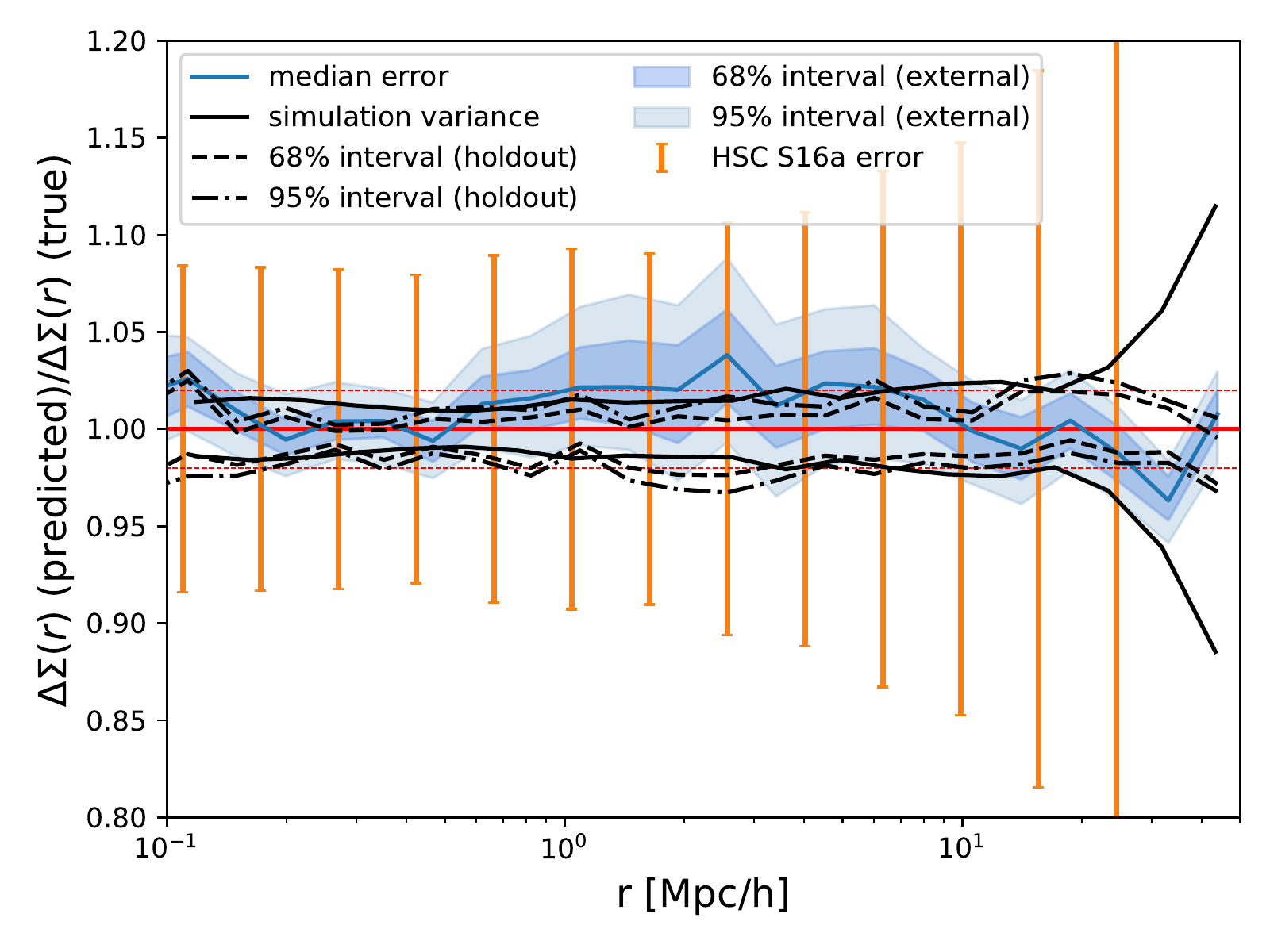}
    \caption{Accuracy test of our $\Delta\Sigma$ emulator. 
    Out-of-sample errors are calculated from the ratio of 150 HODs measured from an external cosmology to the emulator predictions. 
    Holdout tests are performed from the emulators built while excluding five central design cosmologies with their 68\% and 95\% confidence intervals indicated in black dashed and dotted-dashed lines respectively. 
    The solid black lines are the jackknife errors from the simulations used in building the emulator itself.
    The error bars in orange are obtained using CMASS DR12 lenses and HSC imaging from the \texttt{S16A} catalog from jackknife resampling.}
    \label{fig:ds_accuracy_test}
\end{figure}

Figures~\ref{fig:wp_accuracy_test}--~\ref{fig:ds_accuracy_test} show the relative error between the predicted and true ($N$-body measured) quantities for $w_p$, $\xi_0^s$, $\xi_2^s$, and $\Delta\Sigma$ respectively.

The error bars shown in Figures~\ref{fig:wp_accuracy_test},~\ref{fig:xi_mono_accuracy_test} and~\ref{fig:xi_quad_accuracy_test} are the observational errors obtained for $w_p$, $\xi_0^s$ and $\xi_2^s$ measured from the BOSS CMASS DR12 galaxy sample using a jackknife resampling of 400 subregions. 
For Figure~\ref{fig:ds_accuracy_test}, we used the HSC weak lensing measurements of the CMASS galaxy sample as lenses.

We use the $i$-band shape measurements from HSC \texttt{S16A} release with the 
\texttt{frankenz}\footnote{Available at \url{https://github.com/joshspeagle/frankenz}} photo$z$ measurements.
Based on the strategy proposed by \citet{singh2017}, we subtract the lensing signal around 0.5 million random locations from the CMASS $\Delta\Sigma$ profile to achieve unbiased lensing measurements.
We also correct the photo$z$ bias using the calibration sample from the deep COSMOS field. 
This sample has better spectroscopic redshift coverage and high-quality 30 band photo$z$. 
And we estimate the uncertainties of the $\Delta\Sigma$ profiles using the jackknife resampling method with 50 subregions.

We perform these measurements using an updated version of the Python galaxy-galaxy lensing code \texttt{dsigma v2.0}\footnote{Available at \url{https://github.com/johannesulf/dsigma}}.
We refer the interested reader to \citet{speagle2019} and \citet{huang2020} for more details about the HSC lensing measurements.

The solid blue lines in Figures~\ref{fig:wp_accuracy_test}-\ref{fig:ds_accuracy_test} are the median errors from the emulator, and the shaded regions represent the 68\% and 95\% confidence intervals. 
There are several important features to note: all the emulators satisfy the error requirements within the 68\% confidence levels, that is, the errors from the observation data are comparable to the errors in the emulators.  
We have also shown, in solid black, the error on the measurements used to build the emulators from the finite volume of simulations themselves. 
These were estimated by using a jackknife resampling technique by populating the out-of-sample volume (M000) with one of the 150 mock HOD catalogs used for testing and dividing it into 125 subregions.

There are a few instances where the 68\% error distribution exceeds the expected scatter from finite volume effects, e.g. $\xi_0^s$ around 1--8 $h^{-1}$Mpc and $>20\; h^{-1}$Mpc in $\xi_2^s$.
However, there does not appear to be an overall bias in the median prediction except for the quadrupole on large scales because the errors from the in-sample validation test (as discussed in the following section), which is performed over all cosmologies, appears symmetric about unity.
Rather than apply a correction, we have opted to remove scales $>10\; h^{-1}$Mpc from the quadrupole moment for future analyses.

\subsection{Holdout Validation Tests}\label{sec:holdouts}
Holdout (in-sample) tests can only provide an upper bound on the error in our emulators, but can be easily carried out in a variety of ways without having to run additional simulations. 
In this section, we perform a holdout test in which one cosmology or HOD model is excluded from the set and the emulator is rebuilt on the remaining models. 
Generally, the accuracy is poorer for holdout tests, since some models may lie close to the design edge and are now no longer part of the emulator's design space. 
To mitigate this problem, we have chosen to exclude only one of five models that lie close to the center of the design at a time. 
We then use this reduced emulator to predict the missing model for 200 HOD models. 

We show the results of our holdout tests in Figures~\ref{fig:wp_accuracy_test}-\ref{fig:ds_accuracy_test} for each of our emulators. 
We have outlined the 68\% and 95\% regions corresponding to the cosmological holdouts in dashed and dotted-dashed lines respectively.
It is pleasing to note that not only is the spread of models comparable to the out-of-sample test, at many scales, the holdout tests show tighter errors. 
This implies that most of the predictions still satisfy the error requirements even with a reduced design and that the accuracy of the emulators is limited by the cosmological sampling as we should expect and no further improvements can be made on the emulators by increasing the size of the HOD sample.
In fact, if the speed of the computations ever became an issue, the number of HOD models could be reduced such that the confidence regions were of the same size as for the cosmological holdouts. 
Furthermore, our emulator appears to be unbiased with respect to the median prediction. 
The out-of-sample tests in Figures~\ref{fig:xi_mono_accuracy_test}~and~\ref{fig:xi_quad_accuracy_test} seem to indicate a preference for the emulator to underestimate the monopole moment around 1-8 $h^{-1}$Mpc and overestimate the quadrupole at $> 20 \; h^{-1}$Mpc. 
But this is not reproduced in the in-sample validation test, which is varied over the full parameter range, suggesting that the emulator is unbiased through an exploration of the entire design space as would be the case for a likelihood analysis. 
Nonetheless, the median quadrupole prediction on large scales can deviate by as much as 2$\sigma$ so we exclude these regions from further analyses.

\subsection{Error Estimation}
As we have seen from the in-sample tests, there is a non-negligible error associated with the predictions made by our emulators. 
Given that the precision achieved by the emulation technique is approaching the statistical errors of currently available data sets, for robust cosmological constraints, it is necessary to account for these when performing a likelihood analysis. 

Our estimate of the total emulator error, $C_{{\rm emu}, ij}^{\rm tot}$, are as follows: 
\begin{equation}
    C_{ij}^{\rm emu} =  C^{\rm ext}_{ij} + C^{\rm sim}_{ij}, 
    \label{eqn:emu_error}
\end{equation}
where $C^{\rm ext}_{ij}$ is the out-of-sample emulator error and $C^{\rm sim}_{ij}$ represents the scatter in the measurements from the Mira-Titan simulations used to build the emulators themselves.
We cannot simply use the errors from the in-sample validation test to determine the covariance of the emulators because this would overestimate the errors in general as the design shrinks, but models that lie on the edge of the design would require the emulator to perform an extrapolation.

Instead, we have opted to use out-of-sample testing to estimate the error in the emulators with respect to 150 HOD models drawn from an external simulation. 
Although this method does not capture any potential cosmological variation in the covariance, this is the closest approximation to the true error. 
We have confirmed this by repeating our holdout test with only the models at the \emph{center} of the design and checking that this estimate of the error is more comparable with the out-of-sample validation test than the in-sample validation test.
This justifies our use of the fractional error obtained from our out-of-sample test using 150 external models. Then the emulator covariance matrix can be calculated from
\begin{equation}
    C^{\rm ext}_{ij}= \frac{1}{N-1} \sum_{k=1}^N (x^k_i-\bar{x_i})(x^k_j-\bar{x_j}),
    \label{eqn:ext_cov}
\end{equation}
where $x^k_{i} = X^k_{i, {\rm sim}} /X^k_{i, {\rm emu}} - 1$ is the fractional error for $X = w_p, \, \xi_0^s, \, \xi_2^s$ and $\Delta\Sigma$ on the $i$th bin of the $k$th realization and $N = 150$ is the total number of estimates.

\begin{figure*}[t]
    \centering
    \includegraphics[width=\linewidth]{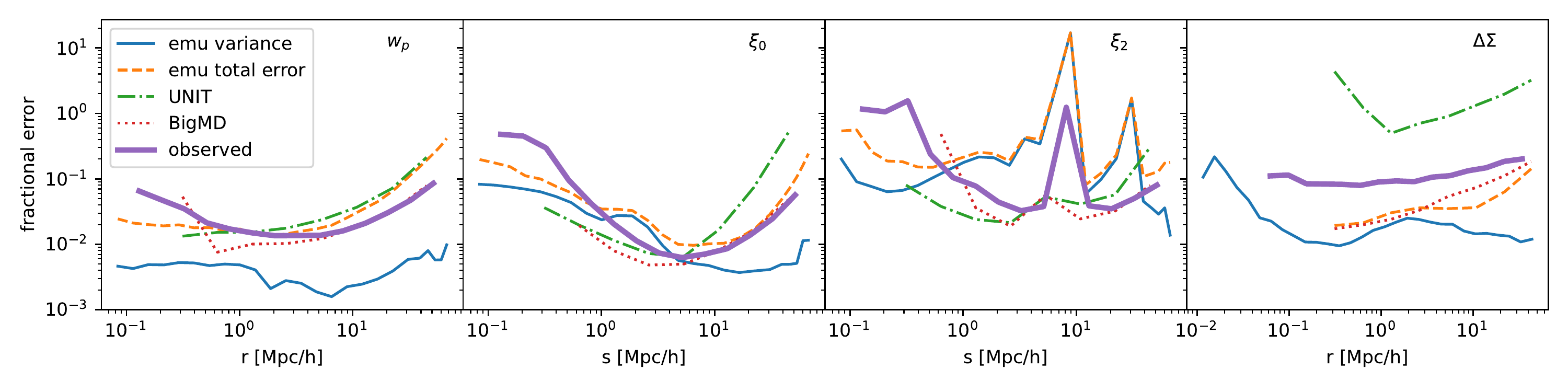}
    \caption{Fractional error for each emulator, defined as $\sigma_X/X$, where $X = w_p, \xi^s_0, \xi^s_2$ and $\Delta\Sigma$, in comparison to the measurement errors. 
    The total emulator error (dashed yellow) is the sum of the emu variance (solid blue) and finite volume error.
    For comparison, we have also shown the relevant observational errors from the CMASS sample (thick purple) and the two mock catalogs we will be using for testing purposes from the UNIT simulation (green dotted-dashed) and BigMultiDark simulation (red dotted) in Section~\ref{sec:results}. 
    }
    \label{fig:emu_error}
\end{figure*}

Additionally, there is a source of error in that the measurements from the Mira-Titan suite themselves contain scatter from various sources such as finite volume effects. 
To determine the effect of these, we use a jackknife resampling technique and break up the simulation volumes into 125 smaller cubes with the measurements repeated as each region is excluded in turn. 
Then the jackknife covariance is formed as follows: 
\begin{equation}
     C^{\rm sim}_{ij} = \frac{N-1}{N} \sum_{k=1}^N (x^k_i-\bar{x_i})(x^k_j-\bar{x_j}),
    \label{eqn:jk_cov}
\end{equation}
with the $x^k_i$'s given by the measurements of $w_p, \, \xi_0^s, \, \xi_2^s$ and $\Delta\Sigma$ themselves (rather than taking their ratios) in the jackknife region $k$ and $N=125$ as the total number of jackknife resamples.
We add the total covariance matrix in Equation~\ref{eqn:emu_error}, encapsulating the theoretical modeling error for all of our emulators to all measurement covariances in the following analyses. 
The sizes of our emulator errors are shown in relation to the observations/mock catalogs that we will consider as test cases in Section~\ref{sec:results} in Figure~\ref{fig:emu_error}. 
We can see that the emulator variance is almost always subdominant to these other errors (except in the case of the quadrupole) and the simulation variance is of approximately the same scale. 
We have confirmed that the above total covariance is a reasonable estimate of the emulation error by applying it to an analysis of mock catalogs formed from the M000 simulation (the out-of-sample model) and verifying that both the correct cosmology and HOD parameters were returned for the full range of scales over $0.1 < s < 50 \, h^{-1}$Mpc.

\section{Application to mock catalogs}~\label{sec:results}
In this section, we test our emulators by applying them to measurements from mock galaxy catalogs with known cosmologies. 
For each probe, we assume a Gaussian likelihood, $\cal{L}$ as follows:
\begin{eqnarray}
    \ln {\cal{L}} \propto -\frac{1}{2} \chi^2, \\
  \chi^2 = ( {\bf x_{M}-x_{D}})^T C^{-1} ({\bf x_{M}-x_{D}}),
\end{eqnarray}
where ${\bf x_{M}}$ and ${\bf x_{D}}$ are the model and data vectors, respectively, and $C$ is the covariance matrix. 
The priors on each parameter are assumed to be uniform over the limits of the emulators for both the HOD and cosmological parameters.

The joint covariances are estimated using a reweighted jackknife resampling method as described in~\citet{mohammed22}, in which we divide the simulation/light-cone into 125 regions. 
For our $\Delta\Sigma$ measurements, we use a smaller region excised out of the full simulation volume to replicate the same signal to noise expected from current survey constraints, since no single WL survey has covered the entire SDSS DR12 footprint. 
Figure~\ref{fig:emu_error} shows that our mock errors approximate the observational CMASS errors well, with the exception of $\Delta\Sigma$ measured from the UNIT mocks, which could be improved by using a larger volume. 
Thus, we expect that our findings in this section will translate to a similar analysis with the observed data vectors. 
We add the emulator errors to this `observational' covariance to form the total covariance matrix. 
Whenever possible, we use the full covariance matrix (including cross-covariances) for the observable quantities and the individual covariance for each emulator. 
Although the cross-covariance between emulators must be nonzero, we expect them to be small. 

The emulators and likelihoods are implemented within the \texttt{CosmoSIS} framework~\citep{zuntz15} for modularity. We use \texttt{Multinest}~\citep{feroz09}, a parallelized Nested Sampling algorithm, to explore the posterior distribution via a Markov Chain Monte Carlo method with 500 live points. 
The sampling terminates when a specified precision criterion has been reached, in this case, the chain is considered converged when the log evidence of the remaining posterior that has yet to be explored can at most add a difference of 0.5 units of the maximum likelihood.  
The final posteriors are visualized using the \texttt{getdist} package~\citep{lewis19}. 

As we show explicitly in the following subsections, we find that our emulators are capable of reproducing the cosmology of these mock catalogs down to $\sim 1\,h^{-1}$Mpc scales. 
This has also been a test of our likelihood analysis pipeline, which will be used in a follow-up publication. 
Our mock pipeline shows that we can measure $\sigma_8$ up to 4.5\% and $f$ up to 7.5\% precision without CMB priors.
With a prior of $\pm 2\sigma$ on $H_0$ from~\citet{planck18}, we can measure $f$ to 2\% precision. 
There are some differences in the HOD preferred parameter space, but these cannot be clearly attributed to a failure of the emulators to describe the clustering in the mocks, rather they could also be due to differences in the definition of halo mass. 

\subsection{UNIT}
The UNIT (Universe $N$-body simulations for the Investigation of Theoretical models) simulations~\citep{chuang19} comprise a set of large-volume, high-resolution simulations run using suppressed variance methods~\citep{pontzen16,angulo16} in which a pair of simulations have initial conditions that have been seeded by the same power spectrum but are out of phase with one another. 
Averaging over measurements from both simulations removes the cosmic variance on large scales.
The volume of the simulations is 1 $h^{-3}$ Gpc$^3$ and contains 4096$^3$ particles, giving a mass resolution of $\sim 1.2 \times 10^9$ $h^{-1}$M$_{\odot}$. 
Halos have been identified down to $\sim 1.2\times 10^{11} h^{-1}$M$_{\odot}$ using \textsc{Rockstar}~\citep{behroozi13}. 
For comparison, the smallest halos contained in Mira-Titan simulations are (on average) $\sim 7.2 \times 10^{11}$ $h^{-1}$M$_{\odot}$ and the smallest allowable value of $\log M_{\rm cut}$ is $\sim 12.3$ in our emulators.  
The cosmology of the  UNIT simulations, taken from~\cite{planck16}, is as follows: $\Omega_m$ = 0.3089, $\Omega_b$ = 0.04086, $h_0$ = 0.6774, $\sigma_8$ = 0.8159, $n_s$ = 0.9667, $w_0$ = -1, $w_a$ = 0 and $\Omega_\nu$ = 0.

\begin{figure*}[t]
    \centering
    \includegraphics[width=\linewidth]{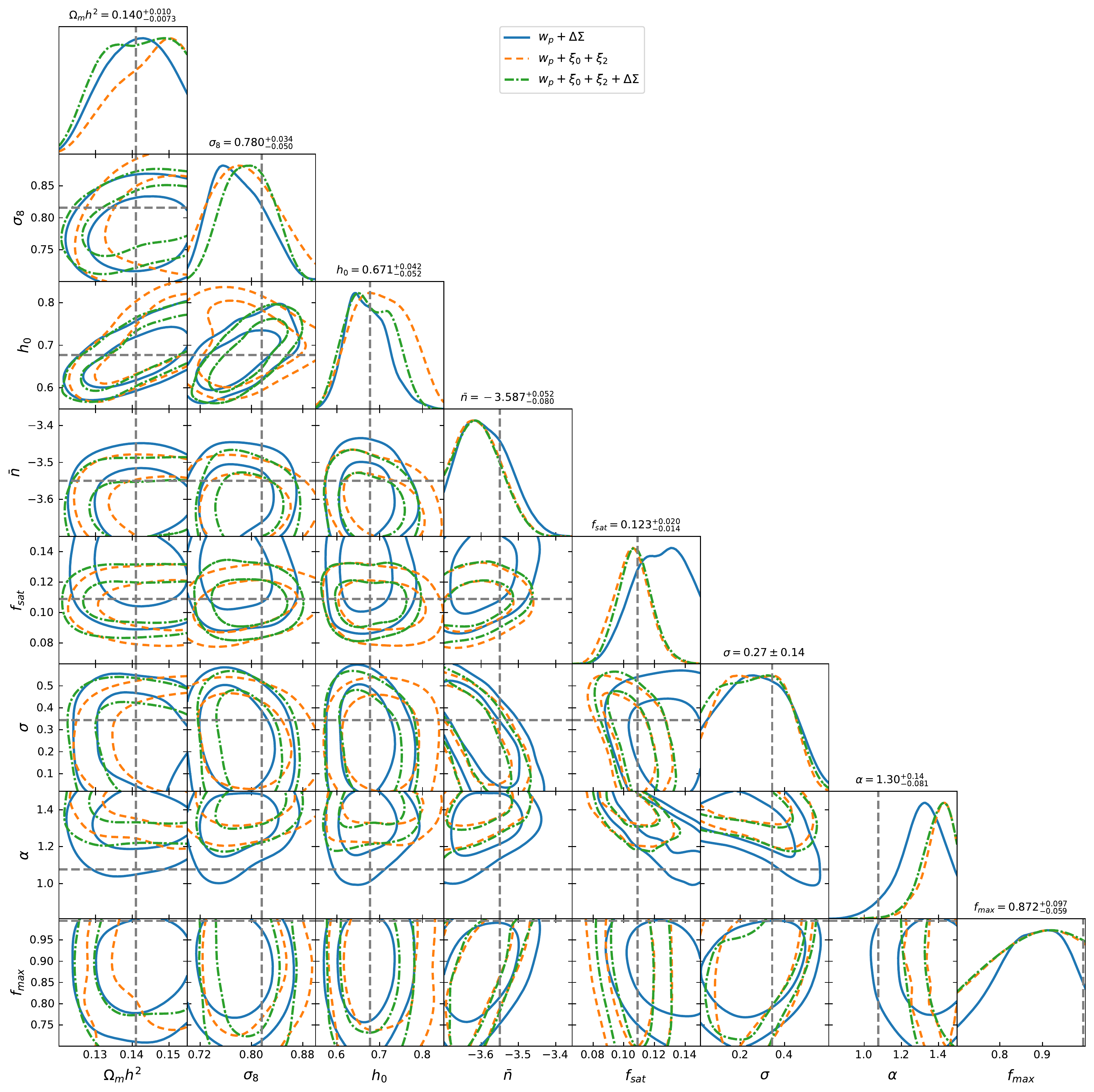}
    \caption{Selected parameter constraints obtained from fitting to the UNIT mock sample with $\sigma_{\rm SHAM} = 0.15$ and $\bar{n}= 2.8 \times 10^{-4} [h^{-1} \rm{Mpc}]^{-3}$ for the following combinations of probes: $w_p + \Delta\Sigma$ (blue solid), $w_p + \xi_0^s + \xi_2^s$ (yellow dashed) and $w_p + \xi_0^s + \xi_2^s +\Delta\Sigma$ (green dotted-dashed). 
    True parameter values are marked by the grey dotted lines, where applicable. 
    This figure shows the posteriors obtained from making a moderate scale cut at 1.25 $h^{-1}$Mpc.
    The parameter ranges are predetermined by the design of our emulators.
    The full triangle plot is contained in Appendix B.}
    \label{fig:unit_contours}
\end{figure*}

\begin{figure*}[t]
    \centering
    \includegraphics[width=\linewidth]{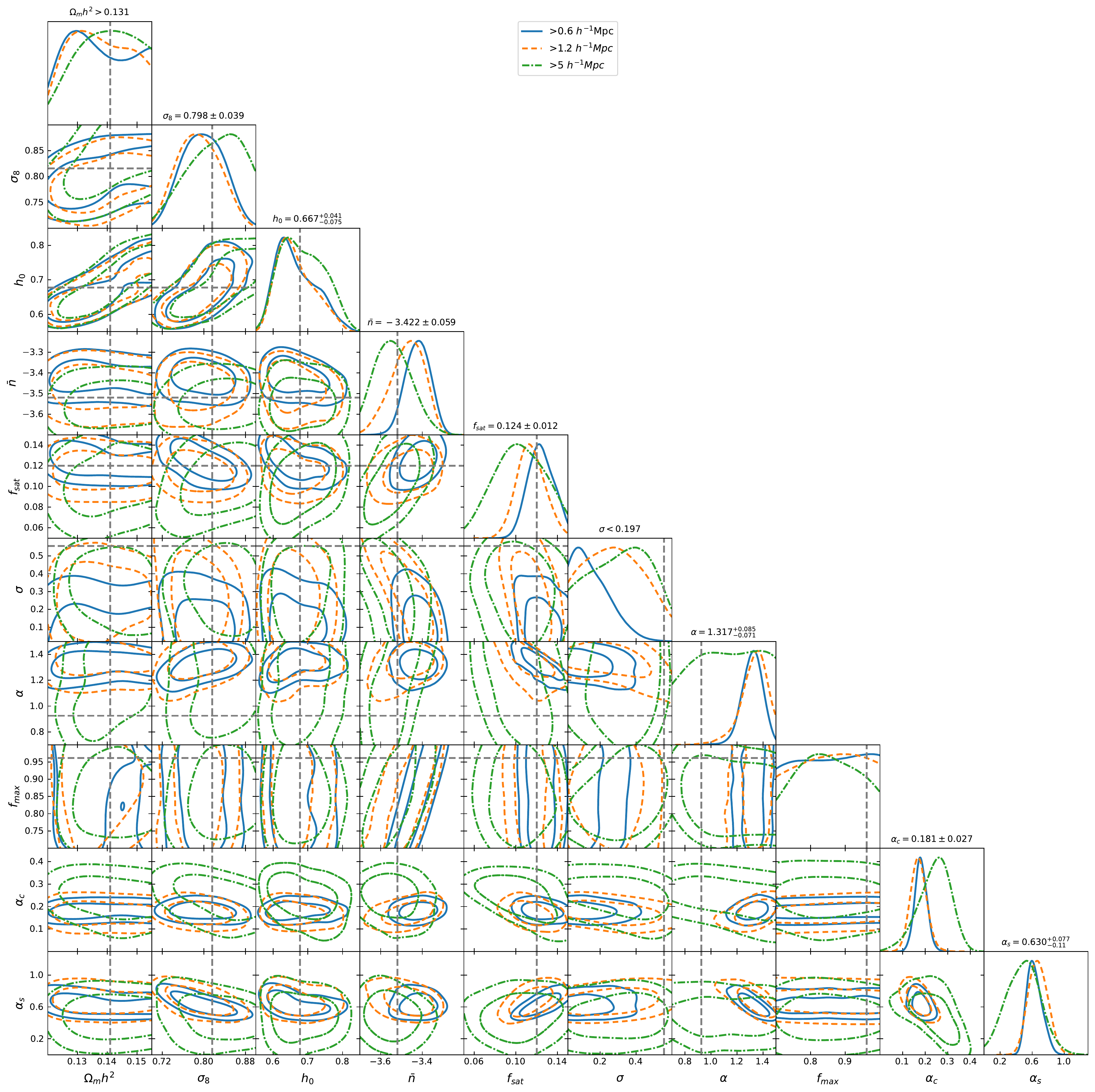}
    \caption{Selected cosmological and HOD constraints obtained from the UNIT mock as a function of discarding different levels of small-scale data for an analysis using $w_p+\xi_0^s+\xi_2^2+\Delta\Sigma$. 
    Solid blue contours denote the use of all scales above 0.6 $h^{-1}$Mpc, dashed yellow contours for a scale cut at 1.2 $h^{-1}$Mpc and dotted-dashed green contours for a scale cut at 5 $h^{-1}$Mpc. Again the triangle plot containing all the parameters that we vary are contained in Appendix B and the parameter ranges correspond to the design limits of our emulators.}
    \label{fig:unit_cosmo_cut}
\end{figure*}

Mock galaxy catalogs were constructed using SHAM from the UNIT simulations using the $z=0.56$ snapshot to match the redshift of the emulators at $z=0.539$. 
We rank order the halos according to their $V_{\rm scat}$ values, defined in the following manner: 
\begin{equation}
    V_{\rm scat} = V_{\rm peak}\left [ 1+N(0,\sigma_{\rm SHAM}) \right],
    \label{eqn:vpeak}
\end{equation}
where $V_{\rm peak}$ is the halo's maximum circular velocity during their lifetime and $\sigma_{\rm SHAM}$
allows for a degree of scatter in the relationship between $V_{\rm peak}$ and stellar mass. 
For this SHAM, we consider two free variables: $\sigma_{\rm SHAM}$ and the number density, $\bar{n}$, of the sample.
When constructing SHAM catalogs, these parameters are constrained by the observed spread in the baryonic Tully-Fisher relation and the measured number density of galaxy sample, but we allow them to vary in this instance in order to test how well our emulators approximate a CMASS-like selection. 
The effect of increasing $\sigma_{\rm SHAM}$ tends to bring in smaller mass halos into the CMASS sample as lower mass halos have a chance to scatter into the sample. 
Meanwhile, increasing $\bar{n}$ obviously lowers the threshold of V$_{\rm scat}$ for entry into the catalog and this decreases the amplitude of clustering as well. 
We have used both $\sigma_{\rm SHAM} = 0.15$ and 0.3 and $\bar{n} = 2.8\times10^{-4} [h^{-1} $Mpc$]^{-3}$ and $3\times10^{-4} [h^{-1} $Mpc$]^{-3}$ as detailed in Table~\ref{tab:unit_results_HOD}; all of which are plausible values for the CMASS sample based on previous SHAM studies~\citep{saito16}. 
From these mocks, we obtained measurements of $w_p, \xi_0^s$, $\xi_2^s$ and $\Delta\Sigma$.

Since many details of these mocks differ from the simple HOD modeling used in the construction of our emulators, e.g., SHAM versus HOD and SO versus FOF halo definitions, it would be unrealistic to expect a perfect replication of all the input parameters (both HOD and cosmological) on all scales. 
This allows us to test some of the assumptions about the extended~\citet{zheng05} HOD model and velocity bias, e.g. that galaxy clustering is solely correlated with halo mass, and how they might impact our constraining power should they be present in the real universe as we explore implementing various cuts in scale to our data vectors. 
We expect that these additional effects to be absent from large scale analyses, say from $> 5\; h^{-1}$Mpc, and so by imposing a number of different values of r$_{\rm min}$ each time, we can determine the level of contamination by comparison. The presence of these additional phenomena beyond the~\citet{zheng05} model is expected to bias the parameter constraints at some level as we gradually relax these scale cuts.
Additionally, we would like to investigate the effect of various measurement combinations, in particular, how much the addition of $\Delta\Sigma$ contributes to the cosmological constraining power.

\begin{deluxetable*}{cccccccc}
\tablenum{1}
    \tablecaption{Cosmological parameters obtained from fitting to UNIT mocks. Rows marked with `mock' show the true values of the simulation. Quoted figures are obtained from the marginalized 1D peak posterior. The value of r$_{\rm min}$ denotes the minimum scale that we use in our analysis. 
    \label{tab:unit_results_cosmo}}
    \tabletypesize{\scriptsize}
    \tablehead{
       \colhead{  }& \colhead{r$_{\rm min}$}& \colhead{$\omega_m$} & \colhead{$\omega_b$} & \colhead{$\sigma_8$} & \colhead{$h$}  & \colhead{$n_s$} & \colhead{f(z=0.534)}\\
       & \colhead{[$h^{-1}$ Mpc]} & & & & & & 
     }
     \startdata
       mock with $\sigma_{\rm SHAM} = 0.15$      & \nodata        &    0.1412    &   0.0223   & 0.8159     & 0.6774        & 0.9667  & 0.766 \\
       $w_p$+$\xi_0$+$\xi_2$         &  1.25     &$0.149^{+0.005}_{-0.011}$&$0.0225^{+0.0006}_{-0.0005}$&$0.776^{+0.058}_{-0.043}$&$0.668^{+0.099}_{-0.036}$&$0.959^{+0.059}_{-0.053}$ & $0.718^{+0.047}_{-0.058}$\\
       $w_p$+$\Delta\Sigma$ &  1.25          &$0.142^{+0.009}_{-0.009}$&$0.0225^{+0.0006}_{-0.0005}$&$0.755^{+0.060}_{-0.024}$&$0.641^{+0.071}_{-0.024}$&$0.982^{+0.038}_{-0.064}$ &$0.713^{+0.105}_{-0.035}$\\
       $w_p$+$\xi_0$+$\xi_2$+$\Delta\Sigma$  &  1.25       
       &$0.149^{+0.0042}_{-0.015}$&$0.0228^{+0.0004}_{-0.0009}$&$0.797^{+0.040}_{-0.035}$&$0.653^{+0.080}_{-0.034}$&$0.936^{+0.078}_{-0.030}$& $0.692^{0.09}_{0.03}$ \\
       $w_p$+$\xi_0$+$\xi_2$+$\Delta\Sigma$  (H$_0$ prior) &  1.25       
       &$0.148^{+0.0046}_{-0.011}$&$0.0231^{+0.0002}_{-0.0009}$&$0.830^{+0.043}_{-0.047}$&$0.669^{+0.010}_{-0.003}$&$0.944^{+0.057}_{-0.057}$& $0.692^{0.09}_{0.03}$ \\
       mock with $\sigma_{\rm SHAM} = 0.3$      & \nodata        &    0.1412    &   0.0223   & 0.8159     & 0.6774        & 0.9667  &0.766\\
       $w_p$+$\xi_0$+$\xi_2$+$\Delta\Sigma$  &  0.6        &$0.130^{+0.021}_{-0.006}$&$0.0220^{+0.0008}_{-0.0004}$&$0.790^{+0.050}_{-0.033}$&$0.633^{+0.074}_{-0.041}$&$0.951^{+0.053}_{-0.048}$ &$0.703^{+0.086}_{-0.040}$\\
       $w_p$+$\xi_0$+$\xi_2$+$\Delta\Sigma$  &  1.25        &$0.132^{+0.016}_{-0.007}$&$0.0223^{+0.0005}_{-0.0006}$&$0.788^{+0.044}_{-0.041}$&$0.638^{+0.064}_{-0.043}$&$0.949^{+0.058}_{-0.051}$ &$0.698^{+0.086}_{-0.035}$\\
       $w_p$+$\xi_0$+$\xi_2$+$\Delta\Sigma$ (H$_0$ prior) &  1.25        &$0.149^{+0.004}_{-0.014}$&$0.0217^{+0.0015}_{-0.0002}$&$0.843^{+0.034}_{-0.058}$&$0.668^{+0.011}_{-0.002}$&$0.950^{+0.050}_{-0.076}$ &$0.748^{+0.028}_{-0.011}$\\
       $w_p$+$\xi_0$+$\xi_2$+$\Delta\Sigma$  &  5        &$0.144^{+0.007}_{-0.013}$&$0.0224^{+0.0004}_{-0.0008}$&$0.852^{+0.033}_{-0.069}$&$0.647^{+0.091}_{-0.042}$&$0.957^{+0.061}_{-0.041}$ &$0.720^{+0.08}_{-0.043}$\\
   \enddata

\end{deluxetable*}

\begin{deluxetable*}{ccccccccc}
\tablenum{2}

    \tablecaption{HOD parameters obtained from fitting to UNIT mocks. The estimated true values of the simulation are shown in the lines labelled with `mock'. Quoted figures are obtained from the marginalized 1D peak posterior. Again, r$_{\rm min}$ refers to the minimum scale used in our analysis.
    \label{tab:unit_results_HOD}}
    \tabletypesize{\scriptsize}
    \tablehead{
       \colhead{  }& \colhead{r$_{\rm min}$}& \colhead{$\bar{n}$} & \colhead{f$_{\rm sat}$} &\colhead{$\sigma$} & \colhead{$\alpha$} & \colhead{$f_{\rm max}$} & \colhead{$\alpha_c$} &\colhead{$\alpha_s$}\\
       & \colhead{[$h^{-1}$ Mpc]} & & & & & & & 
     }
     \startdata
       mock with $\sigma_{\rm SHAM} = 0.15$  &\nodata & -3.55 & 0.109 & 0.344 & 1.076 & 0.995 & \nodata & \nodata \\
       $w_p$+$\xi_0$+$\xi_2$         &  1.25     
       &$-3.61^{+0.07}_{-0.05}$&$0.110^{+0.013}_{-0.012}$&$0.215^{+0.17}_{-0.13}$&$1.40^{+0.075}_{-0.077}$&$0.913^{+0.07}_{-0.11}$&$0.168^{+0.039}_{-0.049}$&$0.778^{+0.075}_{-0.129}$  \\
       $w_p$+$\Delta\Sigma$ &  1.25      
        &$-3.625^{+0.09}_{-0.04}$&$0.131^{+0.012}_{-0.024}$&$0.249^{+0.184}_{-0.128}$&$1.307^{+0.129}_{-0.086}$&$0.924^{+0.044}_{-0.111}$&$0.610^{+0.649}_{-0.267}$&$0.643^{+0.508}_{-0.391}$  \\
       $w_p$+$\xi_0$+$\xi_2$+$\Delta\Sigma$   &  1.25       
       &$-3.612^{+0.059}_{-0.060}$&$0.107^{+0.011}_{-0.011}$&$0.166^{+0.226}_{-0.077}$&$1.434^{+0.054}_{-0.085}$&$0.907^{+0.078}_{-0.102}$&$0.145^{+0.026}_{-0.032}$&$0.706^{+0.111}_{-0.090}$  \\
       $w_p$+$\xi_0$+$\xi_2$+$\Delta\Sigma$ (H$_0$ prior) &  1.25       
       &$-3.605^{+0.050}_{-0.067}$&$0.108^{+0.010}_{-0.011}$&$0.119^{+0.258}_{-0.046}$&$1.438^{+0.052}_{-0.067}$&$0.907^{+0.058}_{-0.129}$&$0.115^{+0.045}_{-0.033}$&$0.680^{+0.100}_{-0.069}$  \\
       mock with $\sigma_{\rm SHAM} = 0.3$ &\nodata & -3.52 & 0.12 & 0.556 & 0.925 & 0.962 & \nodata & \nodata \\
       $w_p$+$\xi_0$+$\xi_2$+$\Delta\Sigma$  &  0.6        
        &$-3.41^{+0.053}_{-0.066}$&$0.122^{+0.013}_{-0.011}$&$0.082^{+0.131}_{-0.058}$&$1.330^{+0.071}_{-0.083}$&$0.950^{+0.019}_{-0.027}$&$0.180^{+0.027}_{-0.024}$&$0.593^{+0.111}_{-0.075}$  \\
       $w_p$+$\xi_0$+$\xi_2$+$\Delta\Sigma$  &  1.25    
        &$-3.466^{+0.089}_{-0.073}$&$0.113^{+0.014}_{-0.013}$&$0.300^{+0.112}_{-0.229}$&$1.356^{+0.086}_{-0.099}$&$0.932^{+0.036}_{-0.156}$&$0.173^{+0.034}_{-0.040}$&$0.657^{+0.132}_{-0.114}$  \\
       $w_p$+$\xi_0$+$\xi_2$+$\Delta\Sigma$ (H$_0$ prior) &  1.25    
        &$-3.467^{+0.098}_{-0.045}$&$0.109^{+0.015}_{-0.009}$&$0.245^{+0.118}_{-0.167}$&$1.373^{+0.088}_{-0.064}$&$0.927^{+0.051}_{-0.169}$&$0.157^{+0.045}_{-0.039}$&$0.592^{+0.120}_{-0.078}$  \\
       $w_p$+$\xi_0$+$\xi_2$+$\Delta\Sigma$  &  5       
        &$-3.55^{+0.086}_{-0.089}$&$0.098^{+0.030}_{-0.018}$&$0.383^{+0.126}_{-0.222}$&$1.376^{+0.052}_{-0.461}$&$0.823^{+0.118}_{-0.066}$&$0.271^{+0.051}_{-0.086}$&$0.551^{+0.147}_{-0.273}$  \\
   \enddata

\end{deluxetable*}

The results of our tests with the UNIT mocks are split into two tables: Table~\ref{tab:unit_results_cosmo} contains the cosmological constraints while Table~\ref{tab:unit_results_HOD} is dedicated to the HOD parameters. 
Note that the growth rate is not a free parameter and is derived from the other parameters fitted during our likelihood analysis. 
We have included it in Table~\ref{tab:unit_results_cosmo} (and Table~\ref{tab:BigMD_results_cosmo}) for comparison with similar studies in the literature. 
The calculation of the corresponding HOD parameters for the SHAM mocks is done with reference to the \textsc{Rockstar} classification of central and satellite galaxies; we simply add up the numbers of centrals and satellites per host halo mass bin and divide by the total to get $\left < N_{\rm cen} \right >$ and $\left < N_{\rm sat} \right >$. 
Then, using the modified~\citet{zheng05} HOD model that we have adopted, we have fitted for the relevant HOD parameters using {\tt scipy's} least squares optimization algorithm. 
The triangle plot of 1 and 2$\sigma$ confidence levels are shown in Figures~\ref{fig:unit_contours} and~\ref{fig:unit_cosmo_cut}. 
We experimented with a number of different small-scale cuts and measurement configurations and found that excluding the clustering measurements below $1.25\;h^{-1}$Mpc is necessary to reproduce the true simulated values of the mocks. 
This is true for all the measurement combinations that we have tested. 
With this limitation, the emulator is capable of reproducing the correct cosmological parameters for all the values of $\sigma_{\rm SHAM}$ and $\bar{n}$ investigated. 
We also found that there was a slight improvement in constraining power over the key cosmological parameters such as $\sigma_8$ and $\omega_m$ when including $\Delta\Sigma$ as a probe, despite the observations of $\Delta\Sigma$ being much lower in signal to noise. 
For example, in the case of $\sigma_{\rm SHAM} = 0.15$ and $\bar{n} = 2.8\times 10^{-4} [h^{-1} $Mpc$]^{-3}$, for scales $>1.2 \; h^{-1}$Mpc, $\sigma_8$ is measured to 4.5\% when using all the probes, compared to 6\% without $\Delta\Sigma$ and 5\% with $w_p$ and $\Delta\Sigma$ alone. 
We have found that for the SHAM configuration of $\sigma_{\rm SHAM}=0.3$ and $\bar{n}$ = $3\times10^{-4} [h^{-1} $Mpc$]^{-3}$ that the emulator is unbiased down to 0.6 $h^{-1}$Mpc in the marginalized cosmological parameters, but some HOD parameters have been grossly misestimated, e.g. the 1D marginalized value for $\sigma$ is $0.082^{+0.131}_{-0.058}$ whereas the actual value is 0.556. 
This is likely due to the Mira-Titan simulations having a lower particle mass resolution than the UNIT mocks and therefore that the emulator is missing the contribution of small halos and halo substructure. 
Furthermore, the cosmological constraints are not substantially degraded if we increase this cutoff to $\sim 5$ $h^{-1}$Mpc, however, the same cannot be said of constraints on the HOD parameters. 
The HOD parameters that become substantially weakened are $f_{\rm sat}, \alpha$, and the velocity bias
parameters.
For example, when employing a more aggressive cutoff such as $>5\;h^{-1}$Mpc compared to $>0.6\;h^{-1}$Mpc, the error on the (mean) value of $\sigma_8$ only increases from 4.9\% to 5.8\%, while the error on $\alpha$ rises from 6\% to 22\%. 
We can conclude that the majority of the additional constraining power in including sub-$h^{-1}$Mpc scales is contributing to the HOD parameters instead of the cosmology. 

Indeed pushing the emulator below sub-$h^{-1}$Mpc causes not only the HOD parameters to be biased, but also the cosmology for $\sigma_{\rm SHAM} = 0.15$ and $\bar{n} = 2.8\times 10^{-4} [h^{-1} $Mpc$]^{-3}$. 
This is not surprising since both $f_{\rm sat}$ and $\alpha$ control the clustering of the satellite galaxies and this dominates the measurement on smaller scales. 
We note that there are substantial differences between the mock catalog produced from the UNIT simulations and the way in which our HOD models are constructed,  including the definition of the halo mass and the assignment of velocities to central galaxies, so we would not expect a good fit in the interior of the halo. 
Although no velocity biases were explicitly incorporated into the construction of the mocks, that is, we did not perturb the subhalo velocities by an additional amount, the velocity bias parameters are always significant (when RSD is involved) regardless of the scales used.
This indicates that there is a difference between the emulators and the catalogs in the intrinsic construction of the mocks themselves, likely due to the velocity assignment.
A small amount of $\alpha_c$ is always preferred indicating that the velocity of the centrals is perturbed away from that of the dark matter, but only slightly. 
The intrahalo velocity term, $\alpha_s$, also makes a significant contribution to the redshift space clustering in the samples that we have considered; our fitted values imply that the satellite galaxies are moving slower than the dark matter content.

\subsection{BigMultiDark}
\citet{rodriguez16} produced mock catalogs of the CMASS DR12 sample by applying a halo
abundance matching technique to the BigMultiDark simulation~\citep{klypin16}. This simulation
has a box length of 2.5 $h^{-1}$Gpc with 3840$^3$ particles and Planck cosmology.  
Mock CMASS galaxies were selected from the simulation by ranking the halos in order of their
peak circular velocity, $V_{\rm peak}$ (with $\sigma_{\rm SHAM}$ = 0.3 in Equation~\ref{eqn:vpeak}) and matching the cumulative number density to the observed stellar mass function, including the measured stellar incompleteness. 
In addition, the light-cone has been produced to have the same redshift distribution and footprint as the full CMASS sample, with weights on each mock galaxy to account for fiber collisions. 
As well as the sample of galaxies, we have also obtained the corresponding light-cone of dark matter particles to enable the calculation of the galaxy-dark matter cross correlation to estimate $\Delta\Sigma$. 

We assume a fiducial Planck cosmology when measuring $w_p$, $\xi_0^s$, $\xi_2^s$ and $\Delta\Sigma$, which allows us to test our implementation of the corrections described in~\cite{more13}. 
These are intended to account for any residual differences between the fiducial cosmology assumed in the measurement of $w_p$ and $\Delta\Sigma$, such as a remapping of angular scales to comoving projected distances, since these two quantities are directly related via the angular diameter distance, which is cosmology dependent. 

The covariances for these mock observations are measured using the same jackknife sampling technique as for the UNIT mocks~\citep{mohammed22}, with the same volume reduction for the $\Delta\Sigma$ measurements in proportion to the coverage of the SDSS footprint by HSC \texttt{S16A}. 
We have also verified by using 500 realizations of the MultiDark-PATCHY simulations~\citep{kitaura16}, a large set of fast and computationally inexpensive mocks based on perturbation theory, that our results are insensitive to the errors in our estimation of the covariance matrix introduced by jackknife resampling, at least to the precision of our emulator and mock volumes. 

\begin{figure*}[t]
    \centering
    \includegraphics[width=\linewidth]{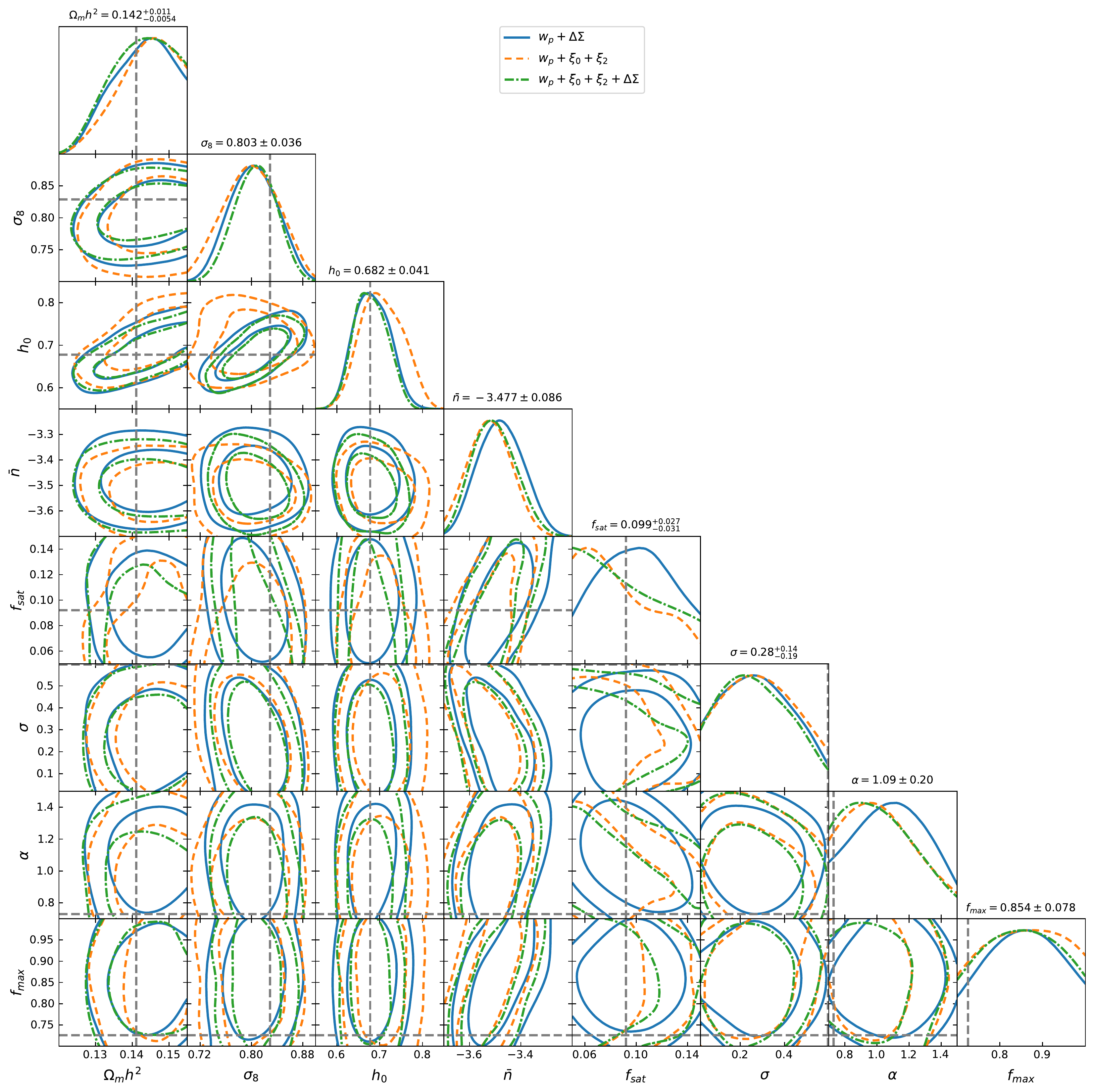}
    \caption{Selected cosmological and HOD constraints obtained from fitting to the BigMultiDark CMASS mock sample for three scenarios, $w_p + \Delta\Sigma$ (blue solid), $w_p + \xi_0^s + \xi_2^s$ (yellow dashed) and $w_p+\xi_0^s+\xi_2^s+\Delta\Sigma$ (green dotted-dashed) using scales above 2.6 $h^{-1}$Mpc. True parameter values are marked by the grey dotted lines. Constraints over the full parameter space are contained in the Appendix and the parameter ranges correspond to the design limits of our emulators.}
    \label{fig:BigMD_cosmo}
\end{figure*}

For the emulators to be able to reproduce the input cosmology, we require a small-scale cutoff of 1.25 $h^{-1}$Mpc for the combination of $w_p+\Delta\Sigma$ and $>2\;h^{-1}$Mpc when all four probes are analyzed together. 
From the values in Tables~\ref{tab:BigMD_results_cosmo} and~\ref{tab:BigMD_results_HOD}, we can see that the combination of $w_p+\xi_0^s+\xi_2^s$ and a cutoff of 1.25 $h^{-1}$Mpc results in values for $h_0$ and $n_s$ that just exceed the true simulation parameters by 1$\sigma$. 
As for the UNIT simulations, the inclusion of $\Delta\Sigma$ to the analysis improves the constraints on $\omega_m$, $h_0$ and $\sigma_8$ even though the $\Delta\Sigma$ measurements have the lowest signal-to-noise. 
When RSD observations are included in $w_p+\Delta\Sigma$ with a cutoff of 1.25 $h^{-1}$Mpc, the value of $n_s$ is unbiased but the lower 1$\sigma$ bound on value of $h_0$ is too high by 4\%.
We can conclude that the problem lies in either the RSD emulation or measurements. 
We note that~\citet{rodriguez16} showed that measurements of $\xi_0^s$ and $\xi_2^s$ from the BigMultiDark lightcone are more than 1$\sigma$ away from the observed values on scales below $\sim h^{-1}$Mpc. 
Furthermore, the effect of fiber collisions (present in the mocks) is not modelled in our emulators; it is our intention to use observations that have been corrected for this effect. 
This has a noticeable effect at $\sim h^{-1}$ Mpc scales for the monopole and up to $10 \, h^{-1}$Mpc for the quadrupole~\citep{rodriguez16}.
Figure~\ref{fig:BigMD_cosmo} shows that the simulation cosmology (contained in Table~\ref{tab:BigMD_results_cosmo}) is recovered in the 1$\sigma$ region of the marginalized posterior distributions when using scales that avoid these problematic regions, i.e. above 2.6 $h^{-1}$Mpc. 

\begin{deluxetable*}{ccccccccc}
\setlength{\extrarowheight}{.5em}
\tablenum{3}
\tablecaption{Cosmological constraints from fitting our emulators to the BigMultiDark CMASS mock. 
The first row represents the true values from the simulation. 
As in the previous tables, we report the marginalized 1D peak posterior and r$_{\rm min}$ is the minimum scale that we used. 
\label{tab:BigMD_results_cosmo}}
    \tablehead{
        \colhead{  }& \colhead{r$_{\rm min}$}& \colhead{$\omega_m$} & \colhead{$\omega_b$} & \colhead{$\sigma_8$} & \colhead{$h$}  & \colhead{$n_s$} &\colhead{f(z=0.533)} \\
        & \colhead{[$h^{-1}$ Mpc]} & & & & & & 
     }
     \startdata
       mock       &\nodata&0.1410& 0.022&  0.8288&0.6777&0.96&0.766\\
        $w_p$ +$\Delta\Sigma$ & 1.25 &$0.147^{+0.005}_{-0.013}$&$0.0227^{+0.0005}_{-0.0007}$&$0.798^{+0.040}_{-0.046}$&$0.667^{+0.069}_{-0.029}$&$0.965^{+0.046}_{-0.047}$&$0.731^{+0.072}_{-0.058}$\\
       $w_p$ +$\Delta\Sigma$ & 2.6 &$0.146^{+0.007}_{-0.009}$&$0.0225^{+0.0006}_{-0.0006}$&$0.812^{+0.026}_{-0.050}$&$0.677^{+0.044}_{-0.041}$&$0.947^{+0.055}_{-0.038}$&$0.722^{+0.088}_{-0.041}$\\
       $w_p$ + $\xi_0^s$ + $\xi_2^s$ & 1.25 &$0.151^{+0.003}_{-0.007}$&$0.0225^{+0.0006}_{-0.0006}$&$0.842^{+0.041}_{-0.042}$&$0.743^{+0.032}_{-0.045}$&$1.027^{+0.020}_{-0.042}$&$0.721^{+0.075}_{-0.046}$\\
       $w_p$ + $\xi_0^s$ + $\xi_2^s$ & 2.6 &$0.145^{+0.008}_{-0.007}$&$0.0220^{+0.0009}_{-0.0003}$&$0.798^{+0.047}_{-0.040}$&$0.690^{+0.063}_{-0.038}$&$0.994^{+0.041}_{-0.054}$ &$0.733^{+0.078}_{-0.052}$\\
       $w_p$+$\xi_0^s$+$\xi_2^s$+$\Delta\Sigma$ & 1.25  &$0.152^{+0.003}_{-0.009}$&$0.0228^{+0.0004}_{-0.0008}$&$0.818^{+0.030}_{-0.035}$&$0.747^{+0.038}_{-0.043}$&$0.994^{+0.042}_{-0.042}$ &$0.735^{+0.067}_{-0.054}$\\
       $w_p$+$\xi_0^s$+$\xi_2^s$+$\Delta\Sigma$ & 2.6  &$0.143^{+0.010}_{-0.007}$&$0.0226^{+0.0004}_{-0.0009}$&$0.807^{+0.034}_{-0.033}$&$0.656^{+0.061}_{-0.020}$&$0.951^{+0.056}_{-0.039}$ &$0.764^{+0.041}_{-0.083}$\\
       $w_p$+$\xi_0^s$+$\xi_2^s$+$\Delta\Sigma$ ($H_0$ prior) & 2.6  &$0.141^{+0.010}_{-0.006}$&$0.0219^{+0.0011}_{-0.0002}$&$0.803^{+0.021}_{-0.026}$&$0.674^{+0.006}_{-0.006}$&$0.965^{+0.043}_{-0.067}$ &$0.753^{+0.027}_{-0.011}$\\
     \enddata

\end{deluxetable*}

\begin{deluxetable*}{ccccccccc}
\setlength{\extrarowheight}{.5em}
\tablenum{4}
\tablecaption{HOD constraints from fitting our emulators to the BigMultiDark CMASS mock. 
As in Table~\ref{tab:unit_results_HOD}, the `true' values of the HOD parameters are fitted from the mocks using our knowledge of which galaxies are centrals or satellites. Values represent the marginalized 1D peak posterior and r$_{\rm min}$ is the minimum scale used. 
\label{tab:BigMD_results_HOD}}
    \tablehead{
       \colhead{  }& \colhead{r$_{\rm min}$}& \colhead{$\bar{n}$} & \colhead{f$_{\rm sat}$} &\colhead{$\sigma$} & \colhead{$\alpha$} & \colhead{$f_{\rm max}$} & \colhead{$\alpha_c$} &\colhead{$\alpha_s$}\\
       & \colhead{[$h^{-1}$ Mpc]} & & & & & & &
     }
     \startdata
        mock      &\nodata&\nodata& 0.092& 0.597&0.729&0.726&\nodata&\nodata\\
        $w_p$ +$\Delta\Sigma$ & 1.25 &$-3.40^{+0.083}_{-0.094}$&$0.125^{+0.018}_{-0.015}$&$0.522^{+0.070}_{-0.230}$&$1.37^{+0.095}_{-0.09}$&$0.854^{+0.097}_{-0.090}$&$0.571^{+0.564}_{-0.357}$&$0.814^{+0.437}_{-0.496}$\\
       $w_p$ +$\Delta\Sigma$ & 2.6 &$-3.49^{+0.096}_{-0.087}$&$0.109^{+0.017}_{-0.041}$&$0.284^{+0.139}_{-0.193}$&$1.058^{+0.260}_{-0.187}$&$0.823^{+0.120}_{-0.056}$&$0.989^{+0.029}_{-0.057}$&$0.839^{+0.397}_{-0.544}$\\
       $w_p$ + $\xi_0^s$ + $\xi_2^s$ & 1.25 &$-3.614^{+0.057}_{-0.068}$&$0.086^{+0.008}_{-0.009}$&$0.567^{+0.032}_{-0.065}$&$1.474^{+0.024}_{-0.074}$&$0.809^{+0.123}_{-0.070}$&$0.117^{+0.027}_{-0.041}$&$0.956^{+0.126}_{-0.087}$\\
       $w_p$ + $\xi_0^s$ + $\xi_2^s$ & 2.6 &$-3.527^{+0.088}_{-0.084}$&$0.065^{+0.036}_{-0.014}$&$0.263^{+0.174}_{-0.163}$&$0.924^{+0.259}_{-0.169}$&$0.916^{+0.049}_{-0.139}$&$0.239^{+0.030}_{-0.091}$&$0.959^{+0.216}_{-0.138}$\\
       $w_p$+$\xi_0^s$+$\xi_2^s$+$\Delta\Sigma$ & 1.25  &$-3.58^{+0.063}_{-0.073}$&$0.08^{+0.009}_{-0.007}$&$0.575^{+0.025}_{-0.054}$&$1.48^{+0.019}_{-0.061}$&$0.750^{+0.148}_{-0.039}$&$0.123^{+0.020}_{-0.018}$&$0.942^{+0.108}_{-0.105}$\\
       $w_p$+$\xi_0^s$+$\xi_2^s$+$\Delta\Sigma$ & 2.6  &$-3.511^{+0.084}_{-0.087}$&$0.0654^{+0.044}_{-0.014}$&$0.223^{+0.177}_{-0.144}$&$0.857^{+0.317}_{-0.107}$&$0.869^{+0.073}_{-0.108}$&$0.221^{+0.032}_{-0.073}$&$0.948^{+0.219}_{-0.141}$\\
       $w_p$+$\xi_0^s$+$\xi_2^s$+$\Delta\Sigma$ ($H_0$ prior) & 2.6  &$-3.489^{+0.078}_{-0.073}$&$0.091^{+0.019}_{-0.014}$&$0.187^{+0.189}_{-0.142}$&$0.846^{+0.298}_{-0.125}$&$0.868^{+0.051}_{-0.135}$&$0.189^{+0.024}_{-0.024}$&$1.007^{+0.141}_{-0.183}$\\
     \enddata

\end{deluxetable*}

\begin{figure*}[t]
    \centering
    \includegraphics[width=\linewidth]{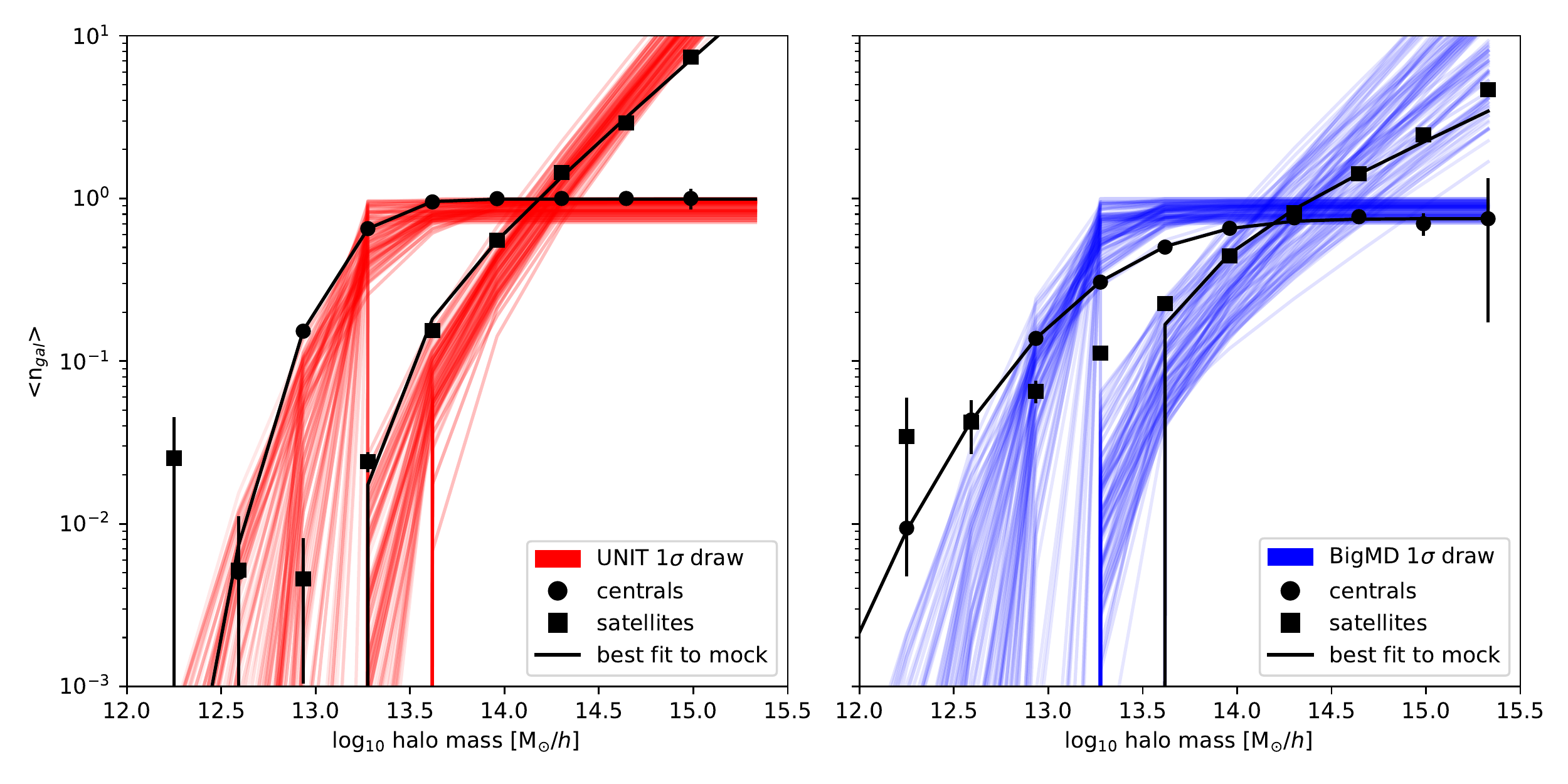}
    \caption{HODs of the UNIT (red; left panel) and BigMultiDark (blue; right panel) mock catalogs. We have measured the HODs directly from the mock galaxy catalogs (with $\left <N_{\rm cen} \right >$ shown as circles and $\left <N_{\rm sat} \right >$ as squares) and compare these to samples drawn from the 1-$\sigma$ distribution of the fitted HOD parameters from a joint analysis of $w_p, \xi_0^s$, $\xi_2^s$ and $\Delta\Sigma$.
    }
    \label{fig:BigMD_HOD}
\end{figure*}

The HOD parameters of the BigMultiDark mock were obtained by measuring the central and satellite fractions in the mocks and fitting them to the~\citet{zheng05} model as described in the previous section for the UNIT mocks. 
Although some of the values, e.g. $\sigma, \alpha, f_{\rm max}$ are close to the edge of the HOD parameter space emulated, the optimization is performed over a much wider range of HOD parameters to account for the possibility that the true HOD is beyond the prior range assumed by our emulator. 
For completeness, the ranges assumed when fitting the HOD are $M_{\rm cut} \in [10.0, 16.0]$, $M_1 \in [10, 16]$, $\sigma \in [0.001, 1.2]$, $\alpha \in [0.001, 1.5]$, and $f_{\rm max} \in [0.01, 1]$. 
Fortunately, the best fitting parameters are within our emulator range and there should be no issue with our emulator modeling the true HOD parameters in this regard. 
Nonetheless, we find that we have mixed success with the HOD parameters, while some, e.g. $\sigma$, can be reproduced, others, e.g. $f_{\rm max}$ and $\alpha$ always exceed the true value by at least 1$\sigma$ and 1.5$\sigma$ respectively. 
The behavior with $\alpha$ is consistent with what we found for the UNIT mocks, namely that the clustering of low-mass halos is missing from the Mira-Titan simulations which is compensated by the inclusion of additional satellite galaxies. 
As in the UNIT mocks, the preferred value of $\alpha_c$ is always slightly greater than zero.   
But in contrast to the UNIT mocks, we find that there is no substantial satellite velocity bias, $\alpha_s$ in the BigMultiDark galaxy catalog, since the 1$\sigma$ contours are consistent with unity.

\section{Discussion}~\label{sec:discussion}
\subsection{Extraction of small-scale information}
The tests in Section~\ref{sec:results} demonstrate that our emulators are capable of analyzing realistic datasets without introducing any biases in the cosmological parameters down to scales of $\sim$ 1$h^{-1}$Mpc in the absence of baryonic effects. 
Fortunately, our investigations with the UNIT mocks show that difference in the constraining power on cosmological parameters is not significantly improved when considering sub-$h^{-1}$ Mpc scales; when we include data down to 0.6 $h^{-1}$ Mpc, we find only a 1\% improvement on the constraint in $\sigma_8$ compared to using 5 $h^{-1}$Mpc as a scale cut (measured from the mean posterior value and 1$\sigma$ uncertainties). 
While there may be some differences in the preferred HOD parameters to the externally measured values, these do not seem to have affected the recovery of the correct cosmology, e.g. in the BigMultiDark mock with a cutoff of $>2\;h^{-1}$Mpc, $\sigma_8$ is (correctly) measured to $\sim$ 4\% precision, but both $\alpha$ and $f_{\rm max}$ are $\sim 1.5\sigma$ from their true values. 
This has two implications: either that the basic extensions of the~\citet{zheng05} model (including the velocity biases) are unable to fully describe the clustering from the SHAM mocks on the sub-$h^{-1}$ Mpc scales and some additional freedom such as allowing the halo concentration to change may be required (as in~\citealt{zhai22} for example) or there is an irreconcilable difference in the halo catalogs since in the Mira-Titan suite the halos are FOF groups while for UNIT and BigMultiDark they are subhalos with SO masses. 
The latter makes a direct comparison more difficult, since the halo masses must necessarily be different which propagates into a shift in the HOD parameters.

Our direct measurements of the HOD parameters in the UNIT and BigMultiDark mock catalogs are shown Figure~\ref{fig:BigMD_HOD} in terms of the mean central and satellite halo occupation as a function of host halo mass. 
In this figure, we compare the best-fitting HOD models (thick colored lines) to the measured points (with centrals represented by circles and squares for satellites) and those sampled from the converged portion of the {\tt Multinest} chain (lightly colored lines). 
As described in the previous Section, the measurement points are derived directly from the central and satellite composition of the mocks and then fitted using {\tt scipy}'s least squares optimization routine.
To convert $\bar{n}$ and f$_{\rm sat}$ into M$_{\rm cut}$ and M$_{1}$ in the {\tt Multinest} chains, we simply swap these values in the original design and rebuild the emulator to return M$_{\rm cut}$ and M$_{1}$ instead. 
Note that this is not a new emulator for the clustering measurements, and that it only predicts M$_{\rm cut}$ and M$_{1}$ from the five original HOD parameters of the extended~\citet{zheng05} model. 
There are two features to note in this figure; first, while the best-fitting HOD values for the mocks might be skewed with regards to what our emulator requires to fit the clustering observations, there is still some overlap in the spread of the allowed 1$\sigma$ region. 
Secondly, the HOD measurement from the mock catalogs contains a lot of scatter for low host halo masses because these are rare objects in the sample. 
This is particularly evident in the $\left< N_{\rm sat} \right>$ counts which have been skewed by the presence of an exceptionally low-mass central with a satellite. 
This results in a poorly fitted HOD model at these low halo masses as well.
Furthermore, these low-mass host halos are below the resolution limit of the Mira-Titan simulations, which makes it particularly difficult for the emulator to describe the contribution of these halos to the clustering signal. 

The UNIT mocks use subhalo clustering information that is not present in the Mira-Titan simulations; this is the most likely cause for the best fitting models to favor higher values of $\alpha$ (i.e. more satellite galaxies) as a remedy for the lack of subhalo information and small-scale clustering.
Furthermore, the mocks have been constructed using SHAMs so it is not surprising that there are some differences with respect to Mira-Titan HODs, especially with in regards to $\alpha$, as the satellite occupation in HOD models is allowed to increase without consideration for mass conservation~\citep{cooray02}, whereas in the case of SHAMs, the physical process of gravitational collapse is required in order for the parent halo to produce more subhalos.

We emphasize that these biases are not seen when analyzing the holdout (M000) mock catalog and that the reproduction of the correct cosmology using measurements on scales down to a few $h^{-1}$Mpc is already quite promising. 
This is encouraging to note since HOD mock galaxy catalogs remain the least computationally expensive in terms of simulation requirements and covering a volume equivalent to the DESI survey with sufficient resolution for a complete subhalo catalog remains challenging.

\subsection{Comparison to other emulators}
There are several other emulators in the literature that also predict galaxy clustering in redshift space and the projected correlation function.
In this section, we provide a comparative discussion of the different approaches.

The \textsc{Aemulus} simulations~\citep{derose19} have spawned the release of several emulators for two-point galaxy clustering statistics, such as~\citet{lange19, zhai19, chapman21, storeyfisher22} and~\citet{zhai22}. 
These simulations cover a similar range in cosmological parameter space as the Mira-Titan simulations, except without the inclusion of massive neutrinos and dynamical dark energy, but do vary the number of relativistic species, $N_{\rm eff}$. 
The cosmological space covered by the \textsc{Aemulus} simulations is built around the~\citet{planck16} best fitting contours. 
While this allows the reduction of the required number of simulations (40 vs. 111 in the Mira-Titan suite), it does apply an implicit prior around the regions of cosmological parameter space favored by Planck, and this is important to note when making claims about consistency between Planck and large-scale structure measurements. 
The \textsc{Aemulus} simulations also have a smaller box size, $L\sim 1.05 \; h^{-1}$Gpc and the lower resolution reduces the available volume for their HOD catalog and parameter space, thus increasing the sample variance in the measurement of their two-point functions and reducing the ability to model low $M_{\rm cut}$ galaxy samples.
The emulators presented in~\citet{zhai19},~\citet{chapman21} and~\citet{zhai22} and are constructed on these simulations and make predictions for the galaxy projected correlation function and redshift space multipole moments as described by the CMASS, LOWZ and eBOSS galaxy samples. 
These emulators use a more complicated modeling of the small-scale clustering, including a concentration parameter to account for baryonic effects and an additional scaling parameter, $\gamma_f$, that operates on all halo velocities to allow for departures from GR. 
The model in \cite{zhai22} also has an additional three parameters describing assembly bias. 
This allows them to include smaller scales down to $\sim~0.1$ $h^{-1}$Mpc in their analysis. 
Both of these studies report a deviation from GR (via a nonzero $\gamma_f$ parameter), at varying degrees of strength (in \citealt{zhai22}, $\gamma_f \ne 1$ is not required for a good $\chi^2$ value) and so merits further study.

An alternative approach using the \textsc{Aemulus} simulation suite is to emulate the Bayesian evidence as developed in~\citet{lange19} and applied to the LOWZ sample in~\citealt{lange21b} and~\citealt{lange23}.
This involves calculating the dependence of the Bayesian evidence as particular target parameters (such as $f\sigma_8$ or S$_8 \equiv \sigma_8 \sqrt{\Omega_m/0.3})$) are varied within the set of $N$-body simulations. 
When calculating the Bayesian evidence, any nuisance parameters, such as those controlling the galaxy-halo connection are marginalized over first for each individual simulation with respect to a particular data set, leaving a single measurement of the target parameters for each cosmology. 
The Bayesian evidence across all simulations in the suite is then approximated with a skewed normal distribution to give a continuous function. 
The advantage of this method is that this technique requires fewer parameters and bypasses any errors associated with the emulation process itself. 
This method is applied in~\citet{lange21b} to obtain a 5\% measurement of $f\sigma_8$ using the monopole, quadrupole and hexadecapole redshift space correlation functions measured from the BOSS LOWZ sample with the theory predictions supplied by the \textsc{Aemulus} simulations. 
When information from galaxy-galaxy lensing is included, ~\citet{lange23} find a similar result to~\citet{chapman21} and~\citet{zhai22}, in that the preferred value of $S_8$ is lower than that reported by Planck.

The Abacus project~\citep{garrison18}, including AbacusCosmos and \textsc{AbacusSummit}~\citep{maksimova21}, are suites of simulations built for the purposes of emulation and used in~\citet{wibking19, wibking20} and~\cite{yuan22a}. 
\citet{wibking19} use an approach similar that in~\citet{zhai19}, except they emulate the projected correlation function and galaxy-galaxy lensing signal from the AbacusCosmos set of simulations.
These simulations are comparable to the \textsc{Aemulus} suite, using 40 $w$CDM cosmologies in [1.1 $h^{-1}$Gpc]$^{3}$ and [720 $h^{-1}$Mpc]$^{3}$ volumes. 
Both $w_p$ and $\Delta\Sigma$ are taken as a ratio with respect to the analytic calculation using the halo model. 
The advantage of this approach is that the emulator for the ratio of quantities is substantially smoother, and that a similar accuracy to~\citet{zhai19} could be achieved with fewer training models. 
These emulators were then applied to the LOWZ sample of galaxies from BOSS~\citep{wibking20}. 
Similarly to~\citet{zhai22} and~\citet{chapman21}, they find a tension between the values of $\Omega_m$  and $\sigma_8$ (actually $S_8$) preferred by their analysis to those predicted by~\citet{planck18} but at a significance of 3.5$\sigma$. 

The emulator presented in \cite{yuan22a} predicts the full 2D redshift space correlation function $\xi(r_p, r_{\pi})$, and is based on the \textsc{AbacusSummit} set of simulations using 85 cosmologies in $(2 h^{-1}$Gpc$)^{3}$ boxes and a mass resolution of $\sim2 \times 10^9 \; h^{-1}$M$_\sun$. 
They also include a number of HOD extensions, such as assembly bias, which enables them to test various HOD prescriptions, and show that their analysis is robust to such changes.
Like many small-scale analyses, e.g.~\citet{zhai22}, ~\citet{yuan22a} report a lower value for the growth rate than~\citet{planck18}, but to a lesser extent than~\citet{zhai22}. 
However, they have opted to not include any extensions to GR. 
The emulator is only trained on the likely regions of cosmological and HOD parameter space -- thus introducing a prior range -- to reduce the amount of simulations required. 
In addition, the \textsc{AbacusSummit} suite does not consider $h$ to be a free parameter, but rather it is derived from the angular diameter distance to the last scattering, $\theta_{\rm rs}$, which is constrained by the CMB to subpercent level for $\Lambda$ cold dark matter ($\Lambda$CDM) models~\citep{planck18}. 
For comparison, in Tables~\ref{tab:unit_results_cosmo} and~\ref{tab:BigMD_results_cosmo} we have also shown the case where a uniform prior on $H_0$ of $\pm 2\sigma$ is applied, which is a loose approximation (because the constraint on $\theta_{\rm rs}$ is much tighter) of the implicit constraint in the \textsc{AbacusSummit} suite. 
This test demonstrates that our emulator can recover the correct growth rate to 2\% precision while using a larger small-scale cutoff than~\citet{yuan22a}.

The Dark Quest simulations~\citep{nishimichi19} are an ensemble of 100 $N$-body simulations in (1 $h^{-1}$Gpc)$^{3}$ volumes that have been used to produce emulators for the halo mass functions and the halo-halo 2D redshift space power spectrum~\citep{kobayashi20}. 
These have been applied to the analysis of $w_p$ and $\Delta\Sigma$ obtained from a combination of LOWZ and CMASS DR12 and HSC \texttt{S16A} observations~\citep{miyatake22a, miyatake22b}. 
As in~\citet{derose19}, this suite covers only massless neutrino cosmologies with a constant dark energy equation of state. 
\citet{kobayashi20} use a different approach to ours; the HOD model is applied afterward analytically, instead of being directly calculated from the $N$-body simulations. 
The galaxy power spectrum is then constructed using an analytic HOD model with nonlinear ingredients supplied by emulators, such as the halo mass function, halo-halo and halo-dark matter power spectra as a function of halo mass. 
This allows for a greater amount of freedom in implementing the galaxy-halo connection, both in choice of HOD model and galaxy sample. 
However, effects that occur on a halo-by-halo basis, such as our implementation of the velocity bias, cannot be modeled in such an approach since these happen between an individual halo and its dark matter particles (or subhalos) and are not encoded in $P^s_{hh}(k,\mu)$ itself. 
Of course this is not an issue if the 1-halo clustering is removed entirely as proposed by~\citet{hikage12}.

In a similar vein, \citet{kokron21} and~\citet{pellejeroibanez23} present emulators that avoids the restrictiveness of HOD modeling by using a mixture of $N$-body simulations and perturbation theory to provide estimates of the galaxy-galaxy and galaxy-dark matter cross power spectra.
The perturbation theory used in these studies is based on a Lagrangian bias expansion presented in~\citet{modi20} and requires calibration to simulation quantities (as supplied by the \textsc{Aemulus} suite) in order to achieve percent-level accuracy. 
The emulator in~\citet{kokron21} has been shown to be $\sim$1\% accurate over the range $\left [ 0.1 \le k \le 1 \right ] h\;$Mpc$^{-1}$ at redshifts $0\le z \le 2$. 
The advantage of their approach lies in its flexibility to model several galaxy populations at once since galaxy biasing is treated via free parameters that are fitted to observations.

Finally, on the subject of comparisons, one important difference that we found, compared to previous works in the literature, was that we were unable to derive stronger constraints on the cosmological parameters from increasing our analysis to smaller scales. 
For example, the error on $\sigma_8$ only reduces by $\sim$1\% when sub-$h^{-1}$Mpc scales are used, whereas the error on $\alpha$ drops by $\sim20$\%. 
This is in contrast to some studies mentioned previously that had advocated for using small-scale information, such as~\citet{wibking19, zhai19, yuan22a}. 
However, it should be noted that we have taken a different approach to these studies when modeling small-scale clustering; we have chosen to use a much simpler model that ignores environmental effects and the halo concentration in favor of having to constrain fewer nuisance parameters. 
We have also made different choices with respect to cosmological parameter space as well, e.g.~\citet{yuan22a} impose a strong $H_0$ constraint. 
In addition, the various emulators presented in the literature also differ in accuracy as a function of scale and thus a careful comparison is required. 
Note that even though these previous works have applied their emulators to observations, the data used are not the same (e.g.~\citealt{lange23} uses the LOWZ galaxy sample and ~\citealt{zhai22} introduce additional color cuts to the CMASS sample).
As there are a number of underlying assumptions that go into building any emulator, it is difficult to clearly differentiate which of these are responsible for the differences between the studies.
It is certainly of interest and importance to the cosmological community to perform a detailed comparison between these emulators to assess the significance of these differences, but that remains beyond the scope of this paper and is left for future investigation.

\subsection{Current limitations and future prospects}
As we approach increasingly accurate measurements of RSD and galaxy-galaxy lensing, it is useful to review the necessary improvements for the next generation of emulators. 
We begin by noting that our current set of emulators is valid only for the CMASS-selected galaxies for which a number of simplifications are justifiable: that the CMASS HOD evolves slowly enough between $0.43 < z < 0.7$ allowing it to be approximated by a single HOD at the median redshift and that the galaxy sample can be well approximated by an HOD (of the~\citealt{zheng05} form) at all.
The former could be remedied by repeating the analysis on other snapshots to extend the HOD model as a function of redshift.

In our approach to modeling galaxy clustering, we have used one of the most basic HOD parameterizations available, however, the presence of assembly bias and baryonic effects could skew the best-fitting cosmological parameters from their true values. 
The inclusion of these effects may allow us to fully utilize observations on sub-megaparsec scales as in~\citet{yuan22a} and~\citet{zhai22}, while the effect of baryons can be mimicked with the baryonification algorithm of \citet{arico21} to a few percent error. 
Indeed, a lack of flexibility in the HOD model may contribute to the so called \emph{lensing-is-low} effect~\citep{chavesmontero23} and the inclusion of environment based assembly bias may be able to mitigate the discrepancy by as much as 50\%~\citep{yuan21a, yuan22a}.

Another consideration for future improvement is a reduction in emulator error and bias; this is especially true for any DESI or LSST analysis where the signal to noise is expected to be much higher. 
Ideally, the errors in the emulator should be $\sim$10\% of the 1-$\sigma$ errors in the data to be considered subdominant. 
To achieve such a stringent criterion would perhaps require the methods of likelihood~\citep{lange19} or ratio emulation~\citep{wibking20, yuan20} to maximise the precision or ultimately a new approach to generating fast predictions from nonlinear scales. 
One disadvantage of our current method is that, while extensions of the HOD and cosmological parameter spaces can be easily incorporated via the Kronecker method, any fundamental changes, say switching to emulating ratios or including assembly bias, will necessitate the building of another emulator, as opposed to simply extending the model with more parameters to capture additional physics, see the various iterations of {\it Halofit} and \texttt{HMCODE}. 

Some of the issues highlighted in this section can be tackled by using a simulation suite of higher resolution and larger volume (thus driving down the intrinsic emulator error) while others may be addressed with a more sophisticated GP application.

\section{Conclusions}
\label{sec:conclusions}
As future large-scale structure surveys increase in sample size and volume, emulation is poised to become a staple of generating theoretical predictions.
We have presented the construction of emulators for the projected correlation function, $w_p$, the redshift space monopole and quadrupole correlation functions, $\xi_0^s \; \xi_2^s$, and the excess surface density, $\Delta\Sigma$ as a function of both cosmological and HOD parameters from the Mira-Titan suite of simulations. 
To reduce the number of nuisance parameters involved, our HOD model is a based on an extension of the classic form presented in~\citet{zheng05}, ~\citet{white11} and~\citet{reid14} with additional parameters for sample incompleteness and velocity biasing. 
We also present a robust set of error covariances associated with the emulators that were estimated using a simulation that was not included in the training set as well as including an extra term for the simulation error using a jackknife resampling of the measurements. 

We have demonstrated that our emulators are sufficiently accurate to analyze currently available observations to constrain both cosmological and HOD parameters on small scales by correctly recovering the input cosmology from several mock galaxy catalogs up to $\sim h^{-1}$Mpc scales. 
These catalogs were made using (S)HAMs to realistically mimic the BOSS CMASS DR12 sample including the effects of the survey geometry and redshift distribution in some cases. 
These findings are robust with respect to variations in the number density and dispersion in the relationship between the stellar mass and the peak halo velocity used to create the SHAM catalog. 
Our results show that there is only a weak improvement in parameter estimation when including scales below $\sim h^{-1}$Mpc in our analysis and that this may actually bias the parameter constraints. 

We will apply the emulators presented here to extract cosmological constraints from the small-scale clustering of the BOSS CMASS sample using a combination of RSD and WL in a future work. This will include a closer study of the baryonic effects on clustering by testing these emulators against mock catalogs made from large-volume fully hydrodynamic $N$-body simulations.

\acknowledgments
We acknowledge the use of the following packages: \texttt{Numpy}\citep{harris20}, \texttt{Scipy}\citep{virtanen20}, \texttt{Matplotlib}~\citep{hunter07}, \textsc{GPy}\footnote{\url{http://github.com/SheffieldML/GPy}}, \texttt{Multinest}\citep{feroz09}, \textsc{CosmoSIS}\citep{zuntz15}, \texttt{getdist}~\citep{lewis19}, \texttt{corrfunc}~\citep{sinha19,sinha20}, \texttt{Halotools}~\citep{hearin17} and \texttt{dsigma v2.0}\footnote{\url{http://github.com/johannesulf/dsigma}}. 

JK acknowledges helpful discussions with Amol Upadhye and Joe DeRose. This project has received funding from the European Research Council (ERC) under the European Union's Horizon 2020 research and innovation programme (grant agreement No 769130).
SH and KH acknowledge support from the U.S. Department of Energy, Office of Science, Office of Advanced Scientific Computing Research and Office of High Energy Physics, Scientific Discovery through Advanced Computing (SciDAC) program under Award Number 231018. 
SS acknowledges the support for this work from NSF-2219212. 
SS is supported in part by World Premier International Research Center Initiative (WPI Initiative), MEXT, Japan.
This material is based on work supported by the U.S. Department of Energy, Office of Science, Office of High Energy Physics under Award Number DE-SC0019301
HG acknowledges the support from the National Science Foundation of China (Nos. 11833005, 11922305). SRT acknowledges partial financial support from
\textit{Ministerio de Ciencia, Innovaci{\'o}n y Universidades / Fondo
Europeo de DEsarrollo Regional} (Spain), under research grant
PGC2018-09497.
The work of the authors at Argonne National Laboratory was supported under the U.S. Department of Energy contract DE-AC02-06CH11357. We acknowledge computational resources made available on the Phoenix compute cluster jointly operated by the Cosmological Physics and Advanced Computing (CPAC) Group and the Computing, Environment and Life Sciences (CELS) Directorate at Argonne National Laboratory. This research used resources of the Argonne Leadership Computing Facility, which is a DOE Office of Science User Facility supported under contract DE-AC02-06CH11357. This research also used resources of the Oak Ridge Leadership Computing Facility, which is a DOE Office of Science User Facility supported under Contract DE-AC05-00OR22725. 
The massive production of all MultiDark-Patchy mocks for the BOSS Final Data Release has been performed at the BSC Marenostrum supercomputer, the Hydra cluster at the Instituto de Fısica Teorica UAM/CSIC, and NERSC at the Lawrence Berkeley National Laboratory. We acknowledge support from the Spanish MICINNs Consolider-Ingenio 2010 Program under grant MultiDark CSD2009-00064, MINECO Centro de Excelencia Severo Ochoa Program under grant SEV- 2012-0249, and grant AYA2014-60641-C2-1-P. The MultiDark-Patchy mocks was an effort led by the IFT UAM-CSIC by F. Prada's group (C.-H. Chuang, S. Rodriguez-Torres and C. Scoccola) in collaboration with C. Zhao (Tsinghua U.), F.-S. Kitaura (AIP), A. Klypin (NMSU), G. Yepes (UAM), and the BOSS galaxy clustering working group.

\section{Appendix A: Emulator Methods for Large Data Sets}
An emulator for both cosmological and HOD parameters necessarily encompasses a large parameter space, with a minimum of five cosmological parameters for the standard $\Lambda$CDM model and another five parameters for a basic HOD model. 
This in turn requires a large number of training models to achieve the required level accuracy for current measurements of small-scale clustering from galaxy surveys. 
Such a large number of training models present a challenge for GPs, since the basic GP scales as $\cal{O}$(N$^3$) as mentioned in Section~\ref{sec:BCM}. 
In this appendix, we detail our exploration of different emulator methods designed to handle large data sets. 

One of the methods we explored was a variant of Bayesian Committee Machines~\citep{tresp00}. 
Our implementation takes advantage of the factorizability of the design into $n_{\rm HOD}\times n_{\rm cosmo}$ models. 
We form a two-tier model of emulation where each cosmology in the Mira-Titan suite is described by a separate GP with its own set of hyperparameters that only accounts for the variation across HOD parameters. 
The output from each emulator forms a new design for another emulator that predicts the desired HOD across the cosmological parameters. 
The final result is an emulator that operates over both HOD and cosmological parameters.

\begin{figure}
    \centering
    \vspace{0.5cm}
    \includegraphics[width=\linewidth]{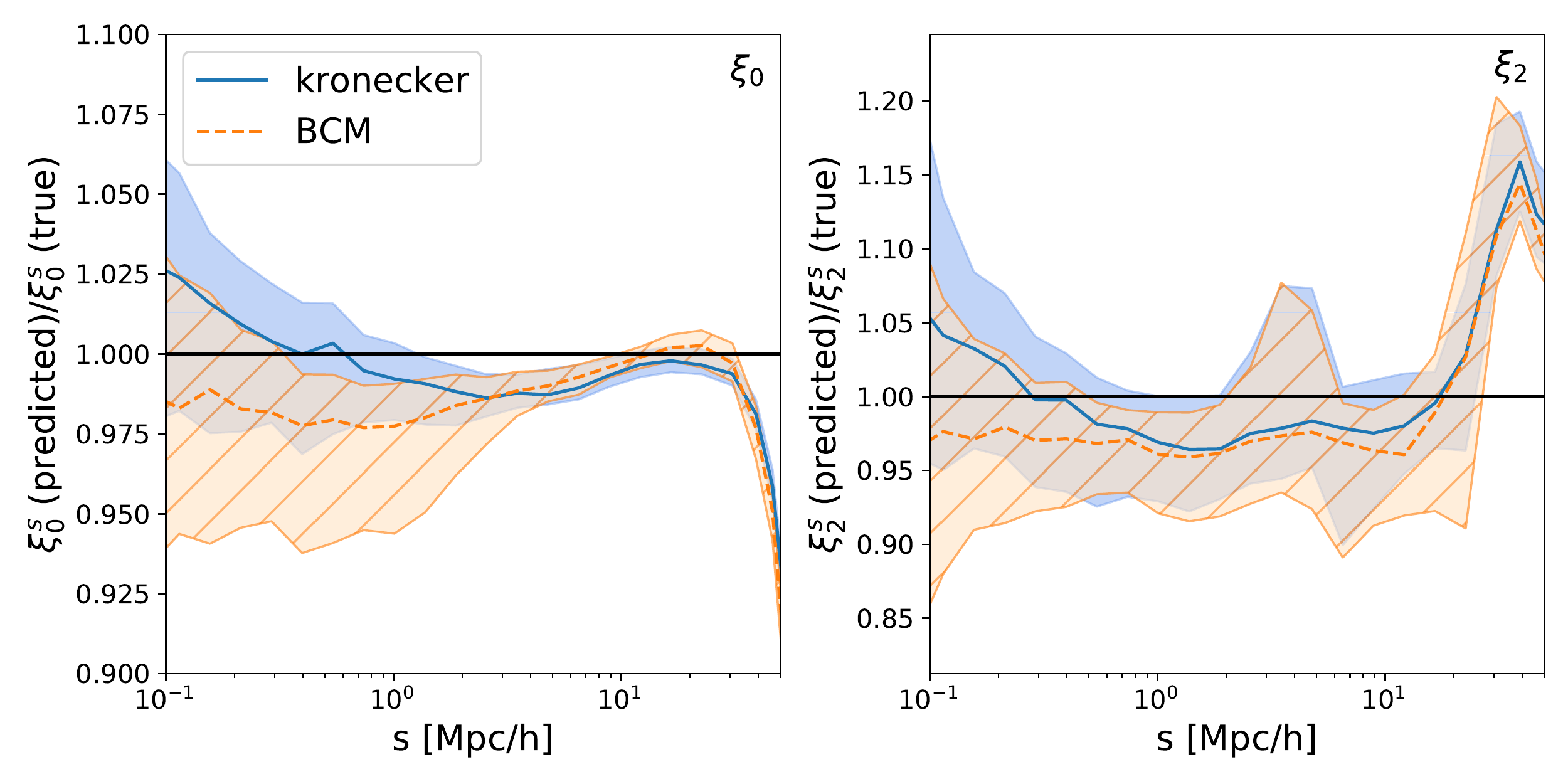}
    \caption{Comparison of the emulators created from a Bayesian Committee Machine (yellow, hatched) and a Kronecker method (blue solid) for the monopole (left) and quadrupole (right).
    The median is shown as a yellow dashed line for the Bayesian Committee Machine and as a blue solid line for the Kronecker method.
    The shaded area is the 68\% confidence interval.}
    \label{fig:AppendixA}
\end{figure}

Figure~\ref{fig:AppendixA} shows the comparison of BCM with the Kronecker method that we used throughout this work. 
We have chosen to show this comparison for only the monopole and quadrupole emulators, since these had the largest emulator errors, one might expect to do better with a different model. 
In general, the BCM method shows more variance and the median error is further away from unity.
This implies that the cosmology and HOD cannot be perfectly factorizable, at least not at the desired precision level because the BCM method treats the cosmology and HOD GPs as separate processes, whereas in the case of the Kronecker GP, only the design is required to be separable.

Note that some of the features remain regardless of the method used, for example the downturn at large-scales in the monopole predictions (which is an upturn for the quadrupole), and the increase in errors near where the quadrupole changes sign.

\section{Appendix B: Full parameter constraints}
For clarity we confine the triangle plots for all the free parameters considered in Figures~\ref{fig:unit_contours}, \ref{fig:unit_cosmo_cut} and \ref{fig:BigMD_cosmo} to this Appendix. 
The setup for each mock test is the same as in Section~\ref{sec:results}, only now the full range of parameters varied is shown.

\begin{figure*}[t]
    \centering
    \includegraphics[width=\linewidth]{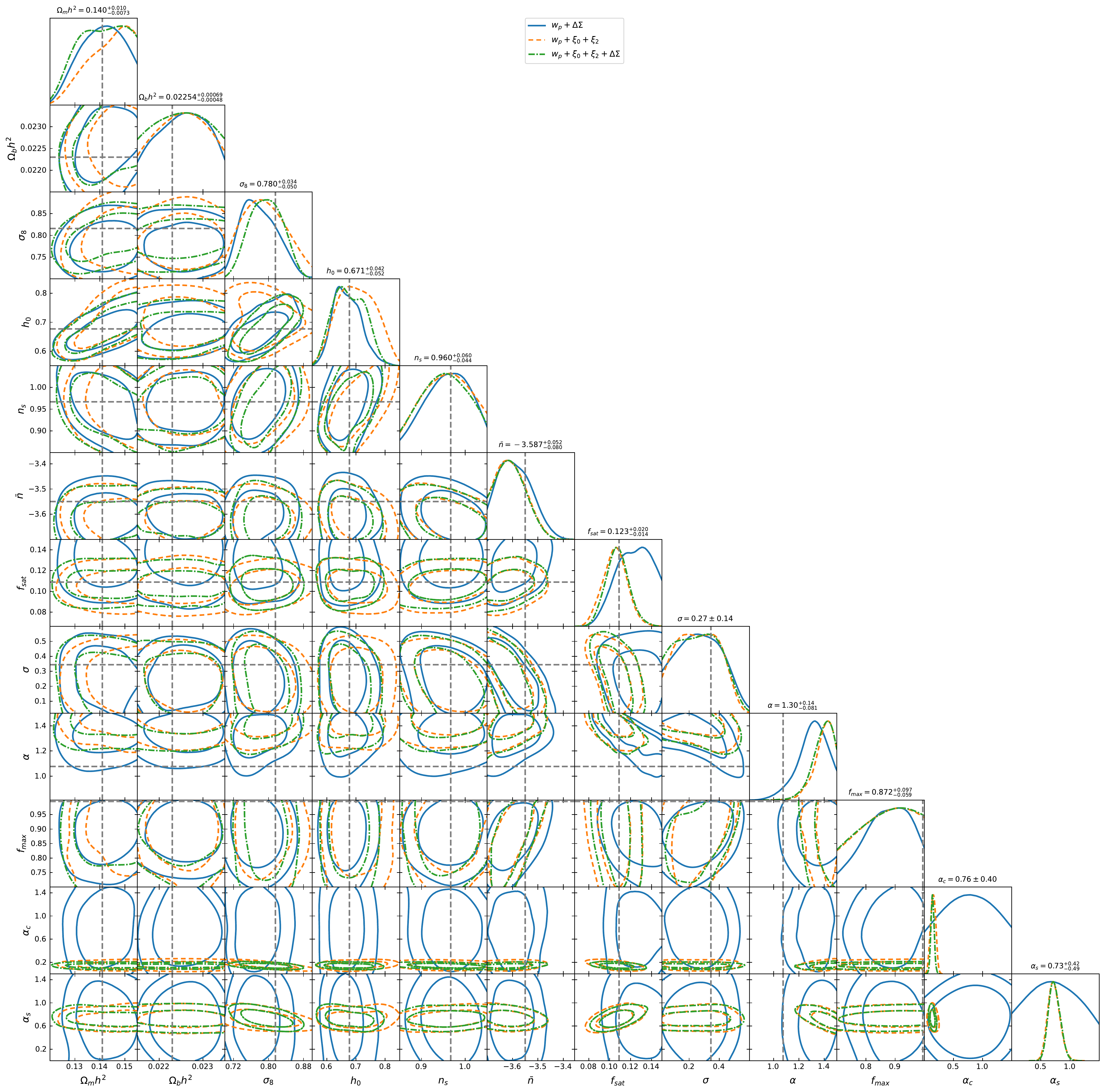}
    \caption{Analog to Figure~\ref{fig:unit_contours} but for all parameters constrained. Contours were obtained from fitting to the UNIT mock sample with $\sigma_{\rm SHAM} = 0.15$ and $\bar{n}= 2.8 \times 10^{-4} [h^{-1} \rm{Mpc}]^{-3}$ for the following combinations of probes: $w_p + \Delta\Sigma$ (blue solid), $w_p + \xi_0^s + \xi_2^s$ (yellow dashed) and $w_p + \xi_0^s + \xi_2^s +\Delta\Sigma$ (green dotted-dashed). 
    True parameter values are marked by the gray dotted lines, where applicable. 
    A moderate scale cut of 1.25 $h^{-1}$Mpc was used for all configurations.
    The parameter ranges are predetermined by the design of our emulators.}
    \label{fig:unit_contours_full}
\end{figure*}

\begin{figure*}[t]
    \centering
    \includegraphics[width=\linewidth]{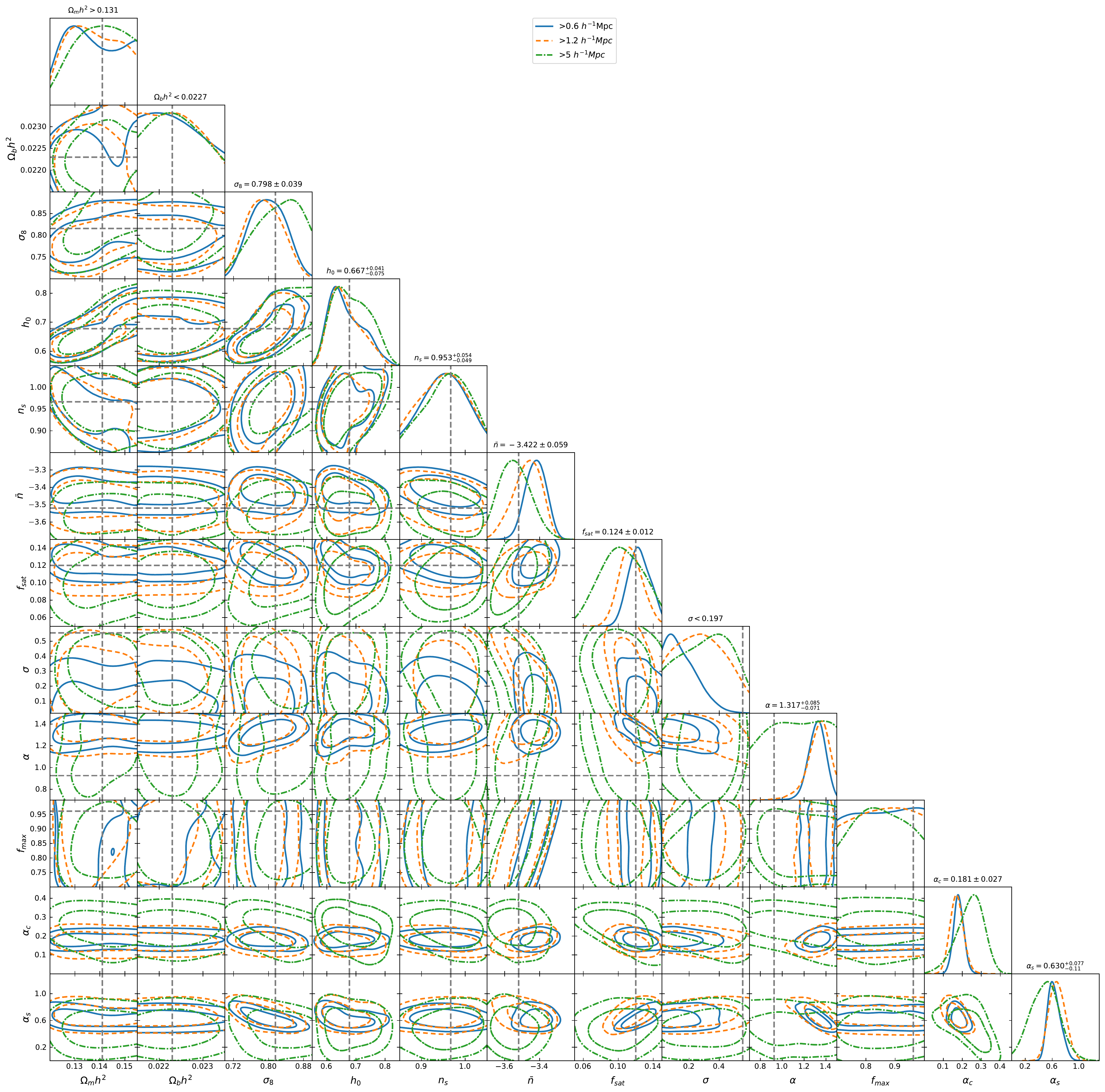}
    \caption{Analog to Figure~\ref{fig:unit_cosmo_cut} but for all free parameters considered in this test. Constraints were obtained from the UNIT mock using the full combination of probes, $w_p + \xi_0^s + \xi_2^s +\Delta\Sigma$, as a function of discarding different levels of small scale data. 
    Blue solid contours denote the use of all scales above 0.6 $h^{-1}$Mpc, yellow dashed contours for a scale cut at 1.2 $h^{-1}$Mpc and green dotted-dashed contours for a scale cut at 5 $h^{-1}$Mpc.}
    \label{fig:unit_cosmo_cut_full}
\end{figure*}

\begin{figure*}[t]
    \centering
    \includegraphics[width=\linewidth]{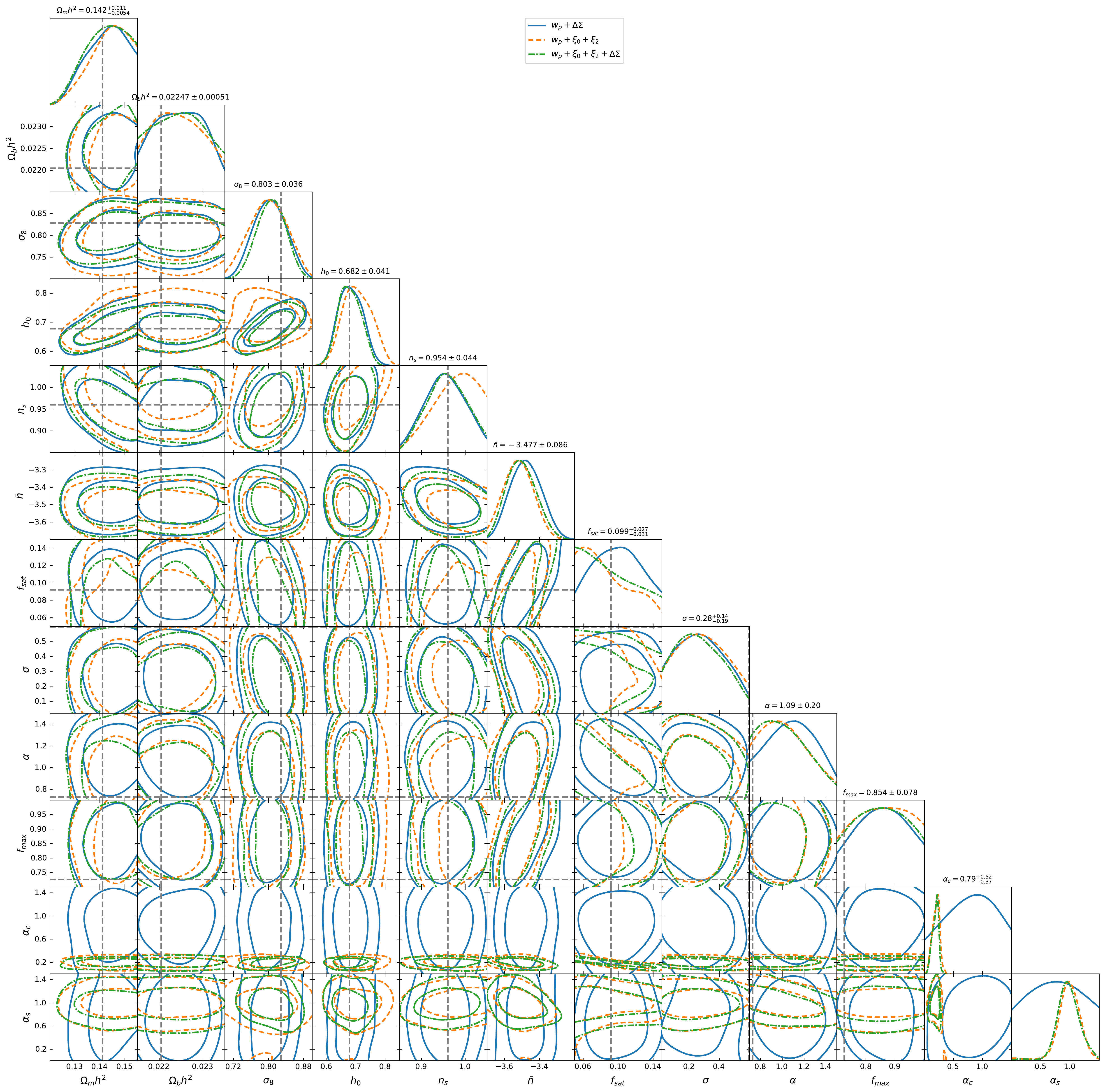}
    \caption{Analog to Figure~\ref{fig:BigMD_cosmo} but for all free parameters considered. Full parameter constraints have been obtained from fitting to the BigMultiDark CMASS mock sample for three scenarios, $\xi_0^s$ and $\xi_2^s$ (blue solid), $w_p + \xi_0^s + \xi_2^s$ (yellow dashed) and $w_p + \xi_0^s+\xi_2^s + \Delta\Sigma$ (green dotted-dashed) using scales above 2.6 $h^{-1}$Mpc. True parameter values are marked by the gray dotted lines.}
    \label{fig:BigMD_cosmo_full}
\end{figure*}

\end{document}